\documentclass[12pt, draftclsnofoot, onecolumn]{IEEEtran}
\usepackage{epsfig,graphicx,subfigure,psfrag,amsmath,cases}
\usepackage{latexsym,amssymb,amsmath,epsfig,subfigure,algorithm,mathtools}
\usepackage{algorithmic}
\usepackage{color}
\usepackage{url}
\usepackage{scrtime}
\usepackage{array}

\author{\authorblockN{Yan Sun, Derrick Wing Kwan Ng, Jun Zhu, and Robert Schober\\
\thanks{Yan Sun, Derrick Wing Kwan Ng, and Robert Schober are  with the Institute for Digital Communications, Friedrich-Alexander-University Erlangen-N\"urnberg (FAU), Germany (email:\{sun, kwan, schober\}@lnt.de). Jun Zhu is with the University of British Columbia, Vancouver, Canada  (email: zhujun@ece.ubc.ca). Derrick Wing Kwan Ng and Robert Schober are also with the University of British Columbia, Vancouver, Canada. This paper has been accepted in part for presentation at the IEEE Globecom 2015 \cite{MOPFD2015confsy}.
}}
}

\title{Multi-Objective Optimization for Robust Power Efficient and Secure Full-Duplex Wireless Communication Systems}

\date{\thistime,\,\today}

\newtheorem{Thm}{Theorem}
\newtheorem{Lem}{Lemma}

\newtheorem{T-Prob}{Transformed Problem}
\newtheorem{E-Prob}{Equivalent Problem}
\newtheorem{SDR-Prob}{SDR Problem}
\newtheorem{Prop}{Proposition}

 \newcommand{\qed}{\hfill \ensuremath{\blacksquare}}

\DeclareMathOperator{\Tr}{\mathrm{Tr}}

\DeclareMathOperator{\Rank}{\mathrm{Rank}}

\DeclareMathOperator{\diag}{\mathrm{diag}}

\DeclareMathOperator{\maxo}{maximize}
\DeclareMathOperator{\mino}{minimize}

\newcommand{\abs}[1]{\lvert#1\rvert}
\newcommand{\norm}[1]{\lVert#1\rVert}

\newcolumntype{L}{>{\arraybackslash\raggedright}m{11cm}}
\makeatletter

\newcommand{\Rmnum}[1]{\expandafter\@slowromancap\romannumeral #1@}
\makeatother

\begin{document}
\maketitle\vspace*{-20mm}
\begin{abstract}
In this paper, we investigate the power efficient resource allocation algorithm design for secure multiuser wireless communication systems employing a full-duplex (FD) base station (BS) for serving multiple half-duplex (HD) downlink (DL) and uplink (UL) users simultaneously. We propose a multi-objective optimization framework to study two conflicting yet desirable design objectives, i.e., total DL transmit power minimization and total UL transmit power minimization. To this end, the weighed Tchebycheff method is adopted to formulate the resource allocation algorithm design as a multi-objective optimization problem (MOOP). The considered MOOP takes into account the quality-of-service (QoS) requirements of all legitimate users for guaranteeing secure DL and UL transmission in the presence of potential eavesdroppers. Thereby, secure UL transmission is enabled by the FD BS and would not be possible with an HD BS. The imperfectness of the channel state information of the eavesdropping channels and the inter-user interference channels is incorporated for robust resource allocation algorithm design. Although the considered MOOP is non-convex, we solve it optimally by semidefinite programming (SDP) relaxation. Simulation results not only unveil the trade-off between the total DL transmit power and the total UL transmit power, but also confirm the robustness of the proposed algorithm against potential eavesdroppers.
\end{abstract}\vspace*{-3mm}
\vspace*{-2mm}
\section{Introduction}
\label{sect1}

The exponential growth in high data rate communication has triggered a tremendous demand for radio resources such as bandwidth and energy. An important technique for reducing the energy and bandwidth consumption of wireless systems while satisfying quality-of-service (QoS) requirements is multiple-input multiple-output (MIMO), as it offers extra degrees of freedom for efficient resource allocation. However, the MIMO gain may be difficult to achieve in practice due to the high computational complexity of MIMO receivers. As an alternative, multiuser MIMO (MU-MIMO) has been proposed as an effective technique for realizing the MIMO performance gain. In particular, in MU-MIMO systems, a transmitter equipped with multiple antennas (e.g. a base station (BS)) serves multiple single-antenna users which shifts the computational complexity from the receivers to the transmitter \cite{MU_MIMO}. Yet, the spectral resource is still underutilized even if MU-MIMO is employed as long as the BS operates in the traditional half-duplex (HD) mode, where uplink (UL) and downlink (DL) communication are separated orthogonally in either time or frequency which leads to a significant loss in spectral efficiency.

Full-duplex (FD) wireless communication has recently received significant attention from both academia and industry due to its potential to double the spectral efficiency of the existing wireless communication systems \cite{FDRad}\nocite{sig_chl_FD,lowcomplex_FDrelay,Dynamic_FDrelay,Mpair_FDRelay_massive}--\cite{FD_smlcll}. In contrast to conventional HD transmission, FD enables simultaneous DL and UL transmission at the same frequency. However, in practice, a major challenge in FD communication is the self-interference (SI) caused by the signal leakage from the DL transmission to the UL signal reception. Although \cite{FDRad}, \cite{sig_chl_FD} reported that SI can be partially cancelled through analog circuits and digital signal processing, the residual SI still severely degrades the performance of FD systems if it is not properly controlled. Besides, co-channel interference (CCI) caused by the UL transmission impairs the DL transmission. Thus, different resource allocation designs for FD systems were proposed  and studied to overcome these challenges. For example, the authors of \cite{lowcomplex_FDrelay} investigated the end-to-end outage probability of MIMO FD single-user relaying systems. In \cite{Dynamic_FDrelay}, a resource allocation algorithm was proposed for the maximization of the end-to-end system data rate of multi-carrier MIMO FD relaying systems. In \cite{Mpair_FDRelay_massive}, massive MIMO was applied in FD relaying systems to facilitate SI suppression and to improve spectral efficiency.  Simultaneous DL and UL transmission via an FD BS in small cells was studied in \cite{FD_smlcll}, where a suboptimal DL beamformer was designed to improve the system throughput.

On the other hand, security is a crucial issue for wireless communication due to the broadcast nature of the wireless medium. Traditionally, secure communication is achieved by cryptographic encryption performed at the application layer and is based on the assumption of limited computational capabilities of the eavesdroppers. However, new computing technologies (e.g. quantum computers) may make this assumption invalid which results in a potential vulnerability of traditional approaches to secure communication. The pioneering work in \cite{wyner1975wire} proposed an alternative approach for providing perfectly secure communication by utilizing the nature of the channel in the physical layer. Specifically, \cite{wyner1975wire} revealed that secure communication can be achieved whenever the information receiver enjoys better channel conditions than the eavesdropper. Inspired by this finding, multiple-antenna transmission has been proposed to ensure communication security \cite{goel2008guaranteeing}\nocite{ng2011secure,liao2011qos}--\cite{li2013spatially}, since multiple antennas provide spatial degrees of freedom which can be utilized to degrade the eavesdropper channel. In particular, transmitting artificial noise (AN) is an effective means to deliberately impair the information reception at the eavesdroppers \cite{goel2008guaranteeing}.
In \cite{ng2011secure}, a power allocation algorithm was designed for maximizing the secrecy outage capacity via AN generation.
The authors of \cite{liao2011qos} proposed a transmit beamforming approach for secrecy provisioning by generating spatially selective AN.
In \cite{li2013spatially}, joint transmit signal and AN covariance matrix optimization was studied for secrecy rate maximization. However, all of the above works focused on HD systems and the obtained results may not be applicable to FD communication systems.
In fact, for FD communication systems, both UL and DL users are exposed to the risk of eavesdropping because of the simultaneous UL and DL transmission. Therefore, it is necessary to ensure communication security for DL and UL concurrently. Although guaranteeing DL security with a multiple-antenna HD BS has been exhaustively studied in the literature \cite{ng2011secure}\nocite{liao2011qos}--\cite{li2013spatially}, UL communication cannot be secured with an HD BS which can perform either transmission or reception in each time instant but not both, and thus, cannot jam the eavesdroppers in the UL. On the other hand, the single-antenna UL users lack the required spatial degrees of freedom to ensure communication secrecy. Multiple-antenna FD BSs are a promising solution to this problem due to their inherent capability of performing simultaneous transmission and reception.

The notion of secure communication in FD systems has received some attention recently. In \cite{zhu2014joint}, joint information beamforming and jamming beamforming for an FD BS was proposed to guarantee DL and UL security. Yet, \cite{zhu2014joint} assumed that there is no CCI between DL and UL users and that the SI at the FD BS can be cancelled perfectly, which may be too optimistic assumptions for practical FD systems. Besides, \cite{zhu2014joint} also assumed that the eavesdropper channel state information (CSI) was perfectly known at the FD BS which which is highly idealistic. In fact, some idle users (e.g. roaming users) in the system may misbehave and eavesdrop the information signal of the legitimate users. Perfect CSI of these potential eavesdroppers may not be available due to their discontinuous interaction with the BS. The authors of \cite{zheng2013improving} proposed an optimal power allocation algorithm to guarantee secure communication for an FD receiver employing only statistical CSI of the eavesdropper channel. In \cite{7015632}, robust beamforming for the case of imperfect CSI was studied for two-way FD communication systems. However, the secure communication approaches proposed in \cite{zheng2013improving} and \cite{7015632} cannot be applied in FD MU-MIMO wireless communication systems directly due to the considerable differences in the considered system models.
Furthermore, to the best of the authors' knowledge, power efficient secure FD communication has not been studied in the literature yet. Specifically, total DL and total UL transmit power minimization are conflicting design objectives in secure FD communication networks. In our previous work \cite{MOPFD2015confsy}, we investigated a power efficient resource allocation algorithm for FD systems under a multi-objective optimization framework which unveiled a trade-off between total DL and total UL power consumption. However, studying this trade-off is still an open problem for secure FD systems. Besides, in \cite{MOPFD2015confsy}, perfect CSI knowledge was assumed.

In this paper, we address the above issues. To this end,  the resource allocation algorithm design for secure FD communication networks is formulated as  a multi-objective optimization problem (MOOP). The proposed MOOP formulation jointly minimizes the total DL transmit power and the total UL transmit power for secure MU-MIMO wireless communication systems employing an FD BS for guaranteeing both UL and DL security. Besides, our problem formulation takes into account the imperfectness of the CSI of the links between the FD BS and the potential eavesdroppers, the links between the UL users and potential eavesdroppers, and the CCI links. Although the considered MOOP is non-convex, we solve it optimally by semidefinite programming (SDP) relaxation leading to a set of Pareto optimal resource allocation policies. Our simulation results not only unveil the trade-off between the total DL transmit power and the total UL transmit power, but also confirm the robustness of the proposed algorithm against imperfect CSI.
\vspace*{-2mm}
\section{System Model}
In this section, we present the considered MU-MIMO FD wireless communication system model.
\vspace*{-4mm}
\subsection{Notation}%
We use boldface capital and lower case letters to denote matrices and vectors, respectively. $\mathbf{A}^H$, $\Tr(\mathbf{A})$, $\Rank(\mathbf{A})$, and $\det(\mathbf{A})$ denotes the Hermitian transpose, trace, rank, and determinant of matrix $\mathbf{A}$, respectively; $\mathbf{A}^{-1}$ and $\mathbf{A}^{\dagger}$ represent the inverse and Moore-Penrose pseudoinverse of matrix $\mathbf{A}$, respectively; $\mathbf{A}\succeq \mathbf{0}$, $\mathbf{A}\succ \mathbf{0}$, and $\mathbf{A}\preceq \mathbf{0}$ indicate that $\mathbf{A}$ is a positive semidefinite, a positive definite, and a negative semidefinite matrix, respectively; $\mathbf{I}_N$ is the $N\times N$ identity matrix; $\mathbb{C}^{N\times M}$ denotes the set of all $N\times M$ matrices with complex entries; $\mathbb{H}^N$ denotes the set of all $N\times N$ Hermitian matrices; $\abs{\cdot}$, $\norm{\cdot}$, and $\norm{\cdot}_F$ denote the absolute value of a complex scalar, the Euclidean vector norm, and the Frobenius matrix norm, respectively; ${\cal E}\{\cdot\}$ denotes statistical expectation; $\diag(x_1, \cdots, x_K)$ denotes a diagonal matrix with the diagonal elements given by $\{x_1, \cdots, x_K\}$ and $\diag(\mathbf{X})$ returns a diagonal matrix having the main diagonal elements of $\mathbf{X}$ on its main diagonal. $\Re(\cdot)$ extracts the real part of a complex-valued input; $[x]^+$ stands for $\mathrm{max}\{0,x\}$; the circularly symmetric complex Gaussian distribution with mean $\mu$ and variance $\sigma^2$ is denoted by ${\cal CN}(\mu,\sigma^2)$; and $\sim$ stands for ``distributed as".
\vspace*{-4mm}

\subsection{Multiuser System Model}
\begin{figure}
\centering\vspace*{-3mm}
\includegraphics[width=4in]{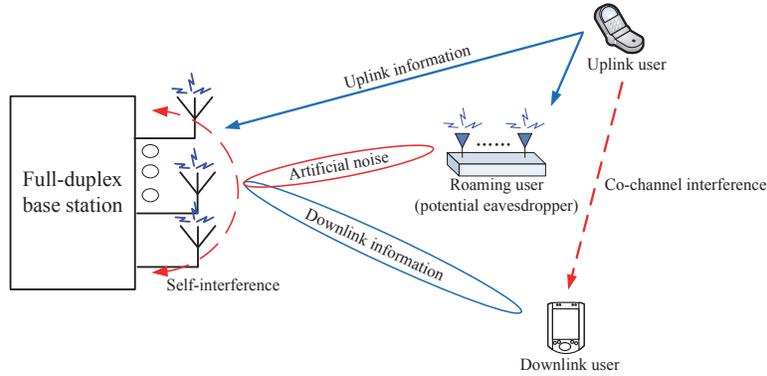}\vspace*{-6mm}
\caption{A multiuser communication system with a full-duplex (FD) radio base station (BS), $K=1$ half-duplex (HD) DL users, $J=1$ HD UL users, and $M=1$ HD roaming user (potential eavesdropper).}
\label{fig:system_model}\vspace*{-8mm}
\end{figure}

We consider a multiuser communication system. The system consists of an FD BS, $K$ legitimate DL users, $J$ legitimate UL users, and $M$ roaming users, cf. Figure \ref{fig:system_model}.
The FD BS is equipped with $N_{\mathrm{T}} > 1$ antennas for facilitating simultaneous DL transmission and UL reception in the same frequency band\footnote{We note that circulator based FD radio prototypes, which can transmit and receive signals simultaneously on the same antennas, have been demonstrated \cite{FDRad}.}. The $K+J$ legitimate users are single-antenna HD mobile communication devices to ensure low hardware complexity. The number of antennas at the FD BS is assumed to be larger than the number of UL users to facilitate reliable UL signal detection, i.e., $N_\mathrm{T} \ge J$. Besides, the DL and the UL users are scheduled for simultaneous UL and DL transmission.
Unlike the local legitimate signal-antenna users, the $M$ roaming users are travelling wireless devices from other communication systems and are equipped with $N_{\mathrm{R}} > 1$ antennas.
The multiple-antenna roaming users are searching for access to local wireless services\footnote{In order to receive the wireless services provided by the local FD BS, the roaming users have to transmit pilot signals to facilitate system clock synchronization and channel estimation.}.
However, it is possible that the roaming users deliberately intercept the information signal intended for the legitimate users if they are in the same service area. As a result, the roaming users are potential eavesdroppers which have to be taken into account for resource allocation algorithm design to guarantee communication security. In this paper, we refer to roaming users as potential eavesdroppers and we assume $N_{\mathrm{T}} > N_{\mathrm{R}}$ for studying resource allocation algorithm design.
\vspace*{-4mm}
\subsection{Channel Model}
We focus on a frequency flat fading channel. In each scheduling time slot, the FD BS transmits $K$ independent signal streams simultaneously at the same frequency to the $K$ DL users. In particular, the information signal to DL user $k \in \{1,\ldots,K\}$ can be expressed as
\begin{eqnarray}
\mathbf{x}_k=\mathbf{w}_k d_k^{\mathrm{DL}},
\end{eqnarray}
where $d_k^{\mathrm{DL}}\in\mathbb{C}$ and $\mathbf{w}_k\in\mathbb{C}^{N_\mathrm{T}\times1}$ are the information bearing signal for DL user $k$ and the corresponding beamforming vector, respectively. Without loss of generality, we assume ${\cal E}\{\abs{d_k^{\mathrm{DL}}}^2\}=1,\forall k\in\{1,\ldots,K\}$.

However, the signal intended for the desired user may be eavesdropped by the roaming users. Hence, in order to ensure secure communication, the FD BS also transmits AN to interfere the reception of the roaming users (potential eavesdroppers). Therefore, the transmit signal vector, $\mathbf{x}\in\mathbb{C}^{N_{\mathrm{T}}\times1}$, comprising $K$ information streams and AN, is given by
\begin{eqnarray}\label{eqn:transmit-signal}
\mathbf{x}=\sum_{k=1}^K \mathbf{x}_k+\mathbf{z},
\end{eqnarray}
where $\mathbf{z}\in\mathbb{C}^{N_\mathrm{T}\times1}$ represents the AN vector generated by the FD BS to degrade the channel quality of potential eavesdroppers.
In particular, $\mathbf{z}$ is modeled as a complex Gaussian random vector with $\mathbf{z}\sim{\cal CN}(\mathbf{0},\mathbf{Z})$,
where $\mathbf{Z}\in\mathbb{H}^{N_\mathrm{T}}$, $\mathbf{Z} \succeq \mathbf{0}$, denotes the covariance matrix of the AN. Therefore, the received signals at DL user $k\in\{1,\ldots,K\}$, the FD BS, and potential eavesdropper $m\in\{1,\ldots,M\}$ are given by \vspace*{-0mm}
\begin{eqnarray}
\label{eqn:dl_user_rcv_signal}y_{k}^{\mathrm{DL}}\hspace*{-2mm} &=&\hspace*{-2mm}\mathbf{h}_k^H\mathbf{x}_k\hspace*{3mm} +\underbrace{\sum_{i \neq k}^K \mathbf{h}_k^H\mathbf{x}_i}_{\mbox{multiuser interference}} + \underbrace{\mathbf{h}_k^H\mathbf{z}}_{\mbox{artificial noise}} +\hspace*{-0.5mm}\underbrace{\sum_{j=1}^J \sqrt{P_j}f_{j,k}d_j^{\mathrm{UL}}}_{\mbox{co-channel interference}}\hspace*{-0.5mm} + \hspace*{3mm} n^{\mathrm{DL}}_{k}\hspace*{-0.5mm},\,\,  \\[0mm]
\label{eqn:ul_rcv_signal}\mathbf{y}^{\mathrm{UL}}\hspace*{-2mm}&=&\hspace*{-2mm}\sum_{j=1}^J \sqrt{P_j}\mathbf{g}_j d_j^{\mathrm{UL}}\hspace*{3mm} + \underbrace{\mathbf{H}_{\mathrm{SI}}\sum_{k=1}^K\mathbf{x}_k}_{\mbox{self-interference}} + \underbrace{\mathbf{H}_{\mathrm{SI}}\mathbf{z}}_{\mbox{artificial noise}} +\hspace*{3mm}\mathbf{n}^{\mathrm{UL}},\,\,\text{and}\\[0mm]
\label{eqn:eve_rcv_signal}\mathbf{y}_m^{\mathrm{E}}\hspace*{-2mm} &=&\hspace*{-2mm}\sum_{k=1}^K\mathbf{L}_m^H\mathbf{x}_k\hspace*{3mm}+\hspace*{3mm}\sum_{j=1}^J \sqrt{P_j}\mathbf{e}_{j,m} d_j^{\mathrm{UL}}\hspace*{3mm} +\underbrace{\mathbf{L}_m^H\mathbf{z}}_{\mbox{artificial noise}} +\hspace*{3mm}\mathbf{n}_m^{\mathrm{E}},\,\,
\end{eqnarray}
respectively. The DL channel between the FD BS and user $k$ is denoted by $\mathbf{h}_k\in\mathbb{C}^{N_{\mathrm{T}}\times1}$ and $f_{j,k}\in\mathbb{C}$ represents the channel between UL user $j$ and DL user $k$. Variables $d_j^{\mathrm{UL}}$, ${\cal E}\{\abs{d_j^{\mathrm{UL}}}^2\}=1$, and $P_j$ are the data and transmit power sent from UL user $j$ to the FD BS, respectively. Vector $\mathbf{g}_j\in\mathbb{C}^{N_{\mathrm{T}}\times1}$ denotes the channel between UL user $j$ and the FD BS. Matrix $\mathbf{H}_{\mathrm{SI}}\in{\mathbb{C}^{N_{\mathrm{T}}\times N_{\mathrm{T}}}}$ denotes the SI channel of the FD BS. Matrix $\mathbf{L}_m \in{\mathbb{C}^{N_{\mathrm{T}}\times N_{\mathrm{R}}}}$ denotes the channel between the FD BS and potential eavesdropper $m$. Vector $\mathbf{e}_{j,m}\in\mathbb{C}^{N_{\mathrm{R}}\times1}$ denotes the channel between UL user $j$ and potential eavesdropper $m$. Variables $\mathbf{h}_k$, $f_{j,k}$, $\mathbf{g}_j$, $\mathbf{H}_{\mathrm{SI}}$, $\mathbf{L}_m$, and $\mathbf{e}_{j,m}$ capture the joint effect of path loss and small scale fading.  $\mathbf{n}^{\mathrm{UL}}\sim{\cal CN}(\mathbf{0},\sigma_{\mathrm{UL}}^2\mathbf{I}_{N_\mathrm{T}})$, $n^{\mathrm{DL}}_{k}\sim{\cal CN}(0,\sigma_{\mathrm{n}_k}^2)$, and $\mathbf{n}_m^{\mathrm{E}}\sim{\cal CN}(\mathbf{0},\sigma_{\mathrm{E}_m}^2\mathbf{I}_{N_\mathrm{R}})$ represent the additive white Gaussian noise (AWGN) at the FD BS, DL user $k$, and potential eavesdropper $m$, respectively.
In (\ref{eqn:dl_user_rcv_signal}), the term $\sum_{j=1}^J \sqrt{P_j}f_{j,k}d_j^{\mathrm{UL}}$ denotes the aggregated CCI caused by the UL users to DL user $k$. In (\ref{eqn:ul_rcv_signal}), the term $\mathbf{H}_{\mathrm{SI}}\sum_{k=1}^K\mathbf{x}_k$ represents the SI.

\section{Resource Allocation Problem Formulation}
In this section, we first define the adopted performance metrics for the considered multiuser communication system. Then, we discuss the assumptions regarding the CSI knowledge for resource allocation. Finally, we formulate the resource allocation problems for DL and UL transmit power minimization, respectively. For the sake of notational simplicity, we define the following variables: $\mathbf{H}_k=\mathbf{h}_k\mathbf{h}_k^H$, $k\in\{1,\ldots,K\}$, and $\mathbf{G}_j=\mathbf{g}_j\mathbf{g}_j^H$, $j\in\{1,\ldots,J\}$.
\vspace*{-2mm}

\subsection{Achievable Rate and Secrecy Rate}
Assuming perfect CSI at the receiver, the achievable rate (bit/s/Hz) of DL user $k$ is given by \vspace*{-0mm}
\begin{eqnarray}
\label{eqn:DL-capacity}R_k^\mathrm{DL} = \log_2(1+\Gamma^{\mathrm{DL}}_{k}), \,\, \mbox{  with}\,\,\,\,
\label{eqn:DL-SINR}\Gamma^{\mathrm{DL}}_{k} = \frac{\abs{\mathbf{h}_k^H\mathbf{w}_k}^2}{\overset{K}{\underset{r \neq k}{\sum}}\abs{\mathbf{h}_k^H\mathbf{w}_r}^2 + \overset{J}{\underset{j=1}{\sum}}P_j\abs{f_{j,k}}^2 + \Tr(\mathbf{H}_k\mathbf{Z}) +\sigma_{\mathrm{n}_k}^2},
\end{eqnarray}
where $\Gamma^{\mathrm{DL}}_{k}$ is the receive signal-to-interference-plus-noise ratio (SINR) at DL user $k$. Besides, the achievable rate of UL user $j$ is given by \vspace*{-2mm}
\begin{eqnarray}
\label{eqn:UL-capacity}R_j^\mathrm{UL}&=&\log_2(1+\Gamma^{\mathrm{UL}}_{j}),\,\, \mbox{  with}\\[-2mm]
\label{eqn:UL-SINR}\Gamma^{\mathrm{UL}}_{j}&=&\frac{P_j\abs{\mathbf{g}_j^H\mathbf{v}_j}^2}{\overset{J}{\underset{n \neq j}{\sum}}P_n\abs{\mathbf{g}_n^H\mathbf{v}_j}^2\hspace*{-0.5mm}+\hspace*{-0.5mm}\Tr\hspace*{-1mm}\Big(\hspace*{-0.5mm}\rho\mathbf{V}_j\diag\Big(\mathbf{H}_{\mathrm{SI}}\mathbf{Z}\mathbf{H}_{\mathrm{SI}}^H \hspace*{-0.5mm}+\hspace*{-0.5mm}\overset{K}{\underset{k=1}{\sum}} \mathbf{H}_{\mathrm{SI}} \mathbf{w}_k\mathbf{w}^H_k\mathbf{H}_{\mathrm{SI}}^H\Big)\Big) \hspace*{-1mm}+\hspace*{-1mm}\sigma_{\mathrm{UL}}^2\norm{\mathbf{v}_j}^2},
\end{eqnarray}
where $\Gamma^{\mathrm{UL}}_{j}$ is the receive SINR of UL user $j$ at the FD BS. The variable $\mathbf{v}_j\in\mathbb{C}^{N_{\mathrm{T}}\times1}$ is the receive beamforming vector for decoding the information received from UL user $j$ and we define $\mathbf{V}_j=\mathbf{v}_j\mathbf{v}_j^H, j\in\{1,\ldots,J\}$. In this paper, zero-forcing receive beamforming (ZF-BF) is adopted. In this context, we note that ZF-BF closely approaches the performance of optimal minimum mean square error beamforming (MMSE-BF) when the noise term is not dominating\footnote{We note that the noise power at the BS is not expected to be the dominating factor for the system performance since BSs are usually equipped with a high quality low-noise amplifier (LNA).} \cite{book:david_wirelss_com} or the number of antennas is sufficiently large \cite{Mpair_FDRelay_massive}. Besides, ZF-BF facilitates the design of a computational efficient resource allocation algorithm.
Hence, the receive beamformer for UL user $j$ is chosen as $\mathbf{v}_j=(\mathbf{u}_j\mathbf{Q}^{\dagger})^H$, where $\mathbf{u}_j=\big[\underbrace{0,\ldots,0}_{(j-1)},1,\underbrace{0,\ldots,0}_{(J-j)}\big]$, $\mathbf{Q}^{\dagger}=(\mathbf{Q}^H\mathbf{Q})^{-1}\mathbf{Q}^H$, and $\mathbf{Q}=[\mathbf{g}_1,\ldots,\mathbf{g}_J]$.
The term $\Tr\Big(\rho\mathbf{V}_j\diag\Big(\mathbf{H}_{\mathrm{SI}}\mathbf{Z}\mathbf{H}_{\mathrm{SI}}^H+\sum_{k=1}^{K} \mathbf{H}_{\mathrm{SI}} \mathbf{w}_k\mathbf{w}^H_k\mathbf{H}_{\mathrm{SI}}^H\Big)\Big)$ in (\ref{eqn:UL-SINR}) models the impact of the imperfectness of the SI cancellation {\cite[Eq. (4)]{JR:FD_model}} due to the limited receiver dynamic range and $0 <\rho \ll 1$ is a constant modelling the noisiness of the SI cancellation at the FD BS. In particular, \cite{JR:ADC} has shown that this model accurately captures the combined effects of additive automatic gain control noise, non-linearities
in the analog-to-digital converters and the gain control, and oscillator phase noise which are present in
practical hardware.

As outlined before, for guaranteeing communication security, roaming users are treated as potential eavesdroppers who eavesdrop the information signals desired for all DL and UL users. Thereby, we design the resource allocation algorithm under a worst-case assumption for guaranteeing communication secrecy. In particular, we assume that a potential eavesdropper can cancel all multiuser interference before decoding the information of a desired user. Thus, under this assumption, the channel capacity between the FD BS and potential eavesdropper $m$ for eavesdropping desired DL user $k$ and the channel capacity between the UL user $j$ and potential eavesdropper $m$ for overhearing UL user $j$ can be written as \vspace*{-1mm}
\begin{eqnarray}
\label{eqn:DL-EVE-rate}C_{k,m}^{\mathrm{DL-E}}&=&\log_2\det(\mathbf{I}_{N_{\mathrm{R}}}+ \mathbf{X}_m^{-1}\mathbf{L}_m^H\mathbf{w}_k\mathbf{w}_k^H\mathbf{L}_m) \,\, \mbox{  and}\\[-0mm]
\label{eqn:UL-EVE-rate}C_{j,m}^{\mathrm{UL-E}}&=&\log_2\det(\mathbf{I}_{N_{\mathrm{R}}}+P_j\mathbf{X}_m^{-1}\mathbf{e}_{j,m}\mathbf{e}_{j,m}^H),
\end{eqnarray}
respectively, where $\mathbf{X}_m=\mathbf{L}_m^H\mathbf{Z}\mathbf{L}_m+\sigma_{\mathrm{E}_m}^2\mathbf{I}_{N_\mathrm{R}}$ denotes the interference-plus-noise covariance matrix for potential eavesdropper $m$. We emphasize that, unlike an HD BS, the FD BS can guarantee both DL security and UL security simultaneously via AN transmission. The achievable secrecy rates between the FD BS and DL user $k$ and UL user $j$ are given by \vspace*{-2mm}
\begin{eqnarray}
\label{eqn:DL-secrecy-rate}R_{k}^{\mathrm{DL-Sec}}&=&\Big[R_k^{\mathrm{DL}}- \underset{m\in\{1,\ldots,M\}}{\mathrm{max}}\big\{C_{k,m}^{\mathrm{DL-E}}\big\}\Big]^+,\,\, \mbox{  and}\\[-2mm]
\label{eqn:UL-secrecy-rate}R_{j}^{\mathrm{UL-Sec}}&=&\Big[R_j^{\mathrm{UL}}- \underset{m\in\{1,\ldots,M\}}{\mathrm{max}}\big\{C_{j,m}^{\mathrm{UL-E}}\big\}\Big]^+,
\end{eqnarray}
respectively.

\vspace*{-4mm}
\subsection{Channel State Information}
In this paper, we focus on slowly time-varying channels. At the beginning of each time slot, the FD BS obtains the CSI of all channels to facilitate global resource allocation.
In practice, the UL users perform handshaking with the FD BS which facilitates UL channel estimation at the FD BS. Since the channels may change slowly in time, the UL users embed pilot signals periodically in the data packets. Hence, the FD BS is able to frequently update and refine the CSI estimate of the UL users. Furthermore, for the acquisition of the CSI of the DL users at the FD BS, handshaking is also performed between the FD BS and the DL users at the beginning of each scheduling slot which allows the FD BS to obtain the statuses and QoS requirements of the DL users. Then during transmission, the DL users are required to send acknowledge (ACK) packets to inform the FD BS of successful reception of the data packets. Hence, the FD BS can regularly update the CSI estimates of the DL transmission links. Therefore, perfect CSI for the UL and DL transmission links, i.e., $\mathbf{g}_j, \forall j \in \{1,\ldots,J\}$, and $\mathbf{h}_k, \forall k \in \{1,\ldots,K\}$, is assumed over the transmission period.
On the other hand, for the CCI channels, the DL users can receive the pilot signals\footnote{We assume that the DL users, UL users, and roaming users utilize orthogonal sequences as pilot signals, which allows the FD BS to distinguish the pilot signals of different users.} of the UL users and feed back the CCI channel estimates to the FD BS only at the beginning of each scheduling time slot. Hence, the FD BS can update the CSI of the CCI channels only at the beginning of every scheduling time slot. As a result, the CSI of the CCI channels at the FD BS is imperfect.
The roaming users (potential eavesdroppers) also perform handshaking at the beginning of a scheduling slot which facilitates the estimation of the corresponding channels at the FD BS. However, the FD BS cannot update the channel information of the potential eavesdroppers during transmission since they are silent in the current time slot. Therefore, only imperfect CSI of the channels between the FD BS and the potential eavesdroppers is available.
Furthermore, for the UL user-to-potential eavesdropper channels, although there is no direct interaction between the potential eavesdroppers and the UL users, the UL users can obtain the CSI by measuring the potential eavesdroppers' pilot signals when the potential eavesdroppers perform handshaking with the FD BS. Then, the UL users can feed back the CSI of these channels to the FD BS. However, since the potential eavesdroppers only perform handshaking at the beginning of each slot, the FD BS can update the information of the UL user-to-potential eavesdropper channels only once in every scheduling time slot. Consequently, only imperfect CSI of the UL user-to-potential eavesdropper channels can be obtained at the FD BS.
To capture the impact of imperfect CSI, we model the CSI uncertainty based on a deterministic model \cite{zheng2008robust}\nocite{jiaheng13Robust}--\cite{wang2009worst}. In particular, the CSI of the link between UL user $j\in\{1,\ldots,J\}$ and DL user $k\in\{1,\ldots,K\}$, i.e., $f_{j,k}$, the CSI of the link between the FD BS and potential eavesdropper $m\in\{1,\ldots,M\}$, i.e., $\mathbf{L}_m$, and the CSI of the link between UL user $j\in\{1,\ldots,J\}$ and potential eavesdropper $m\in\{1,\ldots,M\}$, i.e., $\mathbf{e}_{j,m}$, are modeled as \vspace*{-2mm}
\begin{eqnarray}
\label{eqn:CCI-channel}
f_{j,k}\hspace*{-2mm}&=&\hspace*{-2mm}{\hat f}_{j,k}  + \Delta f_{j,k},\,  \,\,\,
\label{eqn:CCI-set}
{\Omega }_{j,k}\triangleq \Big\{ f_{j,k}\in \mathbb{C}  :\abs{\Delta f_{j,k}} \le \varepsilon_{j,k}\Big\},\\[-2mm]
\label{eqn:BS-EVE-channel}\mathbf{L}_m\hspace*{-2mm}&=&\hspace*{-2mm}\mathbf{\hat L}_m + \Delta\mathbf{L}_m,\,\,\,
\label{eqn:BS-EVE-set}\mathbf{\Omega}_{\mathrm{DL}_m} \triangleq \Big\{\mathbf{L}_m\in \mathbb{C}^{N_{\mathrm{T}}\times N_{\mathrm{R}}}  :\norm{\Delta\mathbf{L}_m}_F \le \varepsilon_{\mathrm{DL}_m}\Big\}, \,\,\, \text{   and}\\[-2mm]
\label{eqn:UL-EVE-channel}\mathbf{e}_{j,m}\hspace*{-2mm}&=&\hspace*{-2mm}\mathbf{\hat e}_{j,m} + \Delta\mathbf{e}_{j,m},\,  \,\,\,
\label{eqn:UL-EVE-set}\mathbf{\Omega}_{\mathrm{UL}_{j,m}} \triangleq \Big\{\mathbf{e}_{j,m}\in \mathbb{C}^{N_{\mathrm{R}}\times 1}  :\norm{\Delta\mathbf{e}_{j,m}} \le \varepsilon_{\mathrm{UL}_{j,m}}\Big\},
\end{eqnarray}
respectively, where ${\hat f}_{j,k}$, $\mathbf{\hat L}_m$, and $\mathbf{\hat e}_{j,m}$ are the CSI estimates available at DL user $k$, the FD BS, and UL user $j$, at the beginning of a scheduling slot, respectively; $\Delta f_{j,k}$, $\Delta\mathbf{L}_m$, and $\Delta\mathbf{e}_{j,m}$ denote the unknown channel uncertainties due to the time varying nature of the channel. The continuous sets ${\Omega }_{j,k}$, $\mathbf{\Omega}_{\mathrm{DL}_m}$, and $\mathbf{\Omega}_{\mathrm{UL}_{j,m}}$ contain all possible channel uncertainties with bounded magnitude $\varepsilon_{j,k}$, $\varepsilon_{\mathrm{DL}_m}$, and $\varepsilon_{\mathrm{UL}_{j,m}}$, respectively. In practice, the values of $\varepsilon_{j,k}$, $\varepsilon_{\mathrm{DL}_m}$, and $\varepsilon_{\mathrm{UL}_{j,m}}$ depend on the coherence time of the associated channels and the transmission duration of the scheduling slot.

\vspace*{-2mm}
\subsection{Optimization Problem Formulation}
We first study the problem formulation for two desirable system design objectives of the considered secure FD communication system. Then, we investigate the two system design objectives jointly under a multi-objective optimization framework. The first considered objective is the minimization of the total DL transmit power at the FD BS and is given by \vspace*{-2mm}
\begin{eqnarray}
\label{eqn:DL-power-min}
&&\hspace*{-10mm}\text{\emph{Problem 1 (Total DL Transmit Power Minimization):}} \notag \\
&&\hspace*{-1mm} \underset{\mathbf{Z}\in\mathbb{H}^{N_\mathrm{T}},\mathbf{w}_k,P_j}{\mino}\,\, \,\, \sum_{k=1}^{K}\norm{\mathbf{w}_k}^2+\Tr(\mathbf{Z}) \notag \\[-3mm]
\notag\mbox{s.t.}
&&\hspace*{-5mm}\mbox{C1: }\underset{\Delta f_{j,k}\in{\Omega}_{j,k}}{\min} \frac{\abs{\mathbf{h}_k^H\mathbf{w}_k}^2}{\overset{K}{\underset{r \neq k}{\sum}}\abs{\mathbf{h}_k^H\mathbf{w}_r}^2 + \overset{J}{\underset{j=1}{\sum}}P_j\abs{f_{j,k}}^2 + \Tr(\mathbf{H}_k\mathbf{Z}) +\sigma_{\mathrm{n}_k}^2} \geq \Gamma^{\mathrm{DL}}_{\mathrm{req}_k},\,\, \forall k, j, \notag\\[-3mm]
&&\hspace*{-5mm}\mbox{C2: }\frac{P_j\abs{\mathbf{g}_j^H\mathbf{v}_j}^2}{\overset{J}{\underset{n \neq j}{\sum}}P_n\abs{\mathbf{g}_n^H\mathbf{v}_j}^2\hspace*{-0.5mm}+\hspace*{-0.5mm}\Tr\hspace*{-1mm}\Big(\hspace*{-0.5mm}\rho\mathbf{V}_j\diag\Big(\mathbf{H}_{\mathrm{SI}}\mathbf{Z}\mathbf{H}_{\mathrm{SI}}^H \hspace*{-0.5mm}+\hspace*{-0.5mm}\overset{K}{\underset{k=1}{\sum}} \mathbf{H}_{\mathrm{SI}} \mathbf{w}_k\mathbf{w}^H_k\mathbf{H}_{\mathrm{SI}}^H\Big)\Big) \hspace*{-1mm}+\hspace*{-1mm}\sigma_{\mathrm{UL}}^2\norm{\mathbf{v}_j}^2} \geq \Gamma^{\mathrm{UL}}_{\mathrm{req}_j},\,\, \forall j,  \notag\\
&&\hspace*{-5mm}\mbox{C3: }\underset{\Delta\mathbf{ L}_m\in \mathbf{\Omega}_{\mathrm{DL}_m}}{\max} \log_2\det(\mathbf{I}_{N_{\mathrm{R}}}+\mathbf{X}_m^{-1}\mathbf{L}_m^H\mathbf{w}_k\mathbf{w}_k^H\mathbf{L}_m) \le R_{\mathrm{tol}_{k,m}}^{\mathrm{DL}} ,\,\, \forall k, m,\notag\\[-2mm]
&&\hspace*{-5mm}\mbox{C4: }\underset{\Delta \mathbf{e}_{j,m} \in \mathbf{\Omega}_{\mathrm{UL}_{j,m}}}{\underset{\Delta\mathbf{ L}_m\in \mathbf{\Omega}_{\mathrm{DL}_m}}{\max}} \log_2\det(\mathbf{I}_{N_{\mathrm{R}}}+P_j\mathbf{X}_m^{-1}\mathbf{e}_{j,m}\mathbf{e}_{j,m}^H) \le R_{\mathrm{tol}_{j,m}}^{\mathrm{UL}} ,\,\, \forall j, m,\notag\\[-0mm]
&&\hspace*{-5mm}\mbox{C5: } P_j \geq 0,\,\, \forall j, \quad
 \mbox{C6: }\mathbf{Z} \succeq \mathbf{0}.
\end{eqnarray}
The system design objective in (\ref{eqn:DL-power-min}) is to minimize the total DL transmit power which is comprised of the DL signal power and the AN power. Constants $\Gamma^{\mathrm{DL}}_{\mathrm{req}_k} > 0$ and $\Gamma^{\mathrm{UL}}_{\mathrm{req}_j} > 0$ in constraints C1 and C2 in (\ref{eqn:DL-power-min}) are the minimum required SINR for DL users $k\in\{1,\ldots,K\}$ and UL users $j\in\{1,\ldots,J\}$, respectively. In particular, in constraint C1, the minimum required SINR for DL user $k$ is satisfied for a given CSI uncertainty set ${\Omega}_{j,k}$ for the CCI channels. $R_{\mathrm{tol}_{k,m}}^{\mathrm{DL}}$ and $R_{\mathrm{tol}_{j,m}}^{\mathrm{UL}}$, in C3 and C4, respectively, are pre-defined system parameters representing the maximum tolerable data rate at potential eavesdropper $m$ for decoding the information of DL user $k$ and UL user $j$, respectively\footnote{If the eavesdroppers do not emit pilot signals, the estimation errors for the eavesdropper channels, $\Delta\mathbf{L}_m$ and $\Delta \mathbf{e}_{j,m}$, are random and follow certain distributions. In this case, the proposed resource allocation algorithm design can still be used but constraints C3 and C4 have to be converted to probabilistic constraints which specify the maximum tolerable secrecy outage probability \cite[Eq. (30), (31)]{li2013spatially}.}. In fact, DL and UL security is guaranteed by constraints C3 and C4 for given CSI uncertainty sets $\mathbf{\Omega}_{\mathrm{DL}_m}$ and $\mathbf{\Omega}_{\mathrm{UL}_{j,m}}$.
In particular, if the above optimization problem is feasible, the proposed problem formulation guarantees that the secrecy rate for DL user $k$ is bounded below by $R_{k}^{\mathrm{DL-Sec}}\ge\log_2(1+\Gamma^{\mathrm{DL}}_{\mathrm{req}_k})-\underset{m}{\mathrm{max}}\{R_{\mathrm{tol}_{k,m}}^{\mathrm{DL}}\}$
and the secrecy rate for UL user $j$ is bounded below by
$R_{j}^{\mathrm{UL-Sec}}\ge\log_2(1+\Gamma^{\mathrm{UL}}_{\mathrm{req}_j})-\underset{m}{\mathrm{max}}\{R_{\mathrm{tol}_{j,m}}^{\mathrm{UL}}\}$. Constraint C5 is the non-negative power constraint for UL user $j$. Constraint C6 and $\mathbf{Z}\in\mathbb{H}^{N_\mathrm{T}}$ are imposed since covariance matrix $\mathbf{Z}$ has to be a Hermitian positive semidefinite matrix. We note that the objective of Problem 1 is to minimize the total DL transmit power under constraints C1--C6 without regard for the consumed UL transmit powers.

The second system design objective is the minimization of total UL transmit power and can be mathematically formulated as\vspace*{-2.5mm}
\begin{eqnarray}
\label{eqn:UL-power-min}
&&\hspace*{-70mm}\text{\emph{Problem 2 (Total UL Transmit Power Minimization):}} \notag \\
&&\hspace*{-15mm} \underset{\mathbf{Z}\in\mathbb{H}^{N_\mathrm{T}},\mathbf{w}_k,P_j}{\mino}\,\, \,\, \sum_{j=1}^{J}P_j \notag\\
\mbox{s.t.} &&\hspace*{-3mm}\mbox{C1 -- C6}.
\end{eqnarray}
Problem 2 targets only the minimization of the total UL transmit power under constraints C1--C6 without taking into account the total consumed DL transmit power.

The objectives of Problems 1 and 2 are desirable for the system operator and the users, respectively. However, in secure FD wireless communication systems, these objectives conflict with each other.
On the one hand, the DL information and AN transmission cause significant SI which impairs the UL signal reception. Hence, the UL users have to transmit with a higher power to compensate this interference to satisfy the minimum required receive SINR of the UL users at the FD BS. On the other hand, a high UL transmit power results in a strong CCI for DL signal reception and a higher risk of information leakage to the potential eavesdroppers. Hence, the FD BS has to transmit both the DL information and the AN with higher power to ensure the QoS requirements of the DL users and the security requirements of the DL and UL users. However, this in turn causes high SI and gives rise to an escalating increase in transmit power for both UL and DL transmission.
To overcome this problem, we resort to multi-objective optimization \cite{JR:MOOP, Kwan15MOOPCR}. In the literature, multi-objective optimization is often adopted to study the trade-off between conflicting system design objectives via the concept of Pareto optimality \cite{JR:MOOP, Kwan15MOOPCR}. To facilitate our presentation, we denote the objective function of Problem $i$ as $Q_i(\mathbf{w}_k,\mathbf{Z},P_j)$. The Pareto optimality of a resource allocation policy is defined in the following:

\emph{Definition  \cite{JR:MOOP}:} A resource allocation policy, $\{\mathbf{w}_k,\mathbf{Z},P_j\}$, is Pareto optimal if and only if there does not exist any $\{\tilde{\mathbf{w}}_k,\tilde{\mathbf{Z}},\tilde{P_j}\}$ with $Q_i(\tilde{\mathbf{w}}_k,\tilde{\mathbf{Z}},\tilde{P_j}) < Q_i(\mathbf{w}_k,\mathbf{Z},P_j)$, $\forall i \in \{1,2\}$.

In other words, a resource allocation policy is Pareto optimal if there is no other policy that improves at least one of the objectives without detriment to the other objective. In order to capture the complete Pareto optimal set, we formulate a third optimization problem to investigate the trade-off between Problem $1$ and Problem $2$ by using the weighted Tchebycheff method \cite{JR:MOOP}. The third problem formulation is given as\vspace*{-2mm}
\begin{eqnarray}
\label{eqn:MOP}
&&\hspace*{-48mm}\text{\emph{Problem 3 (Multi-Objective Optimization):}} \notag \\
&&\hspace*{-10mm}\underset{\mathbf{Z}\in\mathbb{H}^{N_\mathrm{T}},\mathbf{w}_k,P_j}{\mino}\,\,\max_{i=1,2}\,\, \Big\{\lambda_i \Big(Q_i(\mathbf{w}_k,\mathbf{Z},P_j)-Q_i^*\Big)\Big\}\nonumber\\
&&\hspace*{6mm}\mbox{s.t.}\hspace*{4mm} \mbox{C1 -- C6},
\end{eqnarray}
where $Q_1(\mathbf{w}_k,\mathbf{Z},P_j)=\sum_{k=1}^{K}\norm{\mathbf{w}_k}^2+\Tr(\mathbf{Z})$ and $Q_2(\mathbf{w}_k,\mathbf{Z},P_j)=\sum_{j=1}^{J}P_j$.
$Q_i^*$ is the optimal objective value of the $i$-th problem and is treated as a constant for Problem 3. Variable $\lambda_i\ge 0$, $\sum_i\lambda_i=1$, specifies the priority of the $i$-th objective compared to the other objectives and reflects the preference of the system operator. By varying $\lambda_i$, we can obtain a complete Pareto optimal set which corresponds to a set of resource allocation policies. Thus, the operator can select a proper resource allocation policy from the set of available policies. Compared to other formulation methods for handing MOOPs in the literature (e.g. the weighted product method and the exponentially weighted criterion \cite{JR:MOOP}), the weighted Tchebycheff method can achieve the complete Pareto optimal set with a lower computational complexity, despite the non-convexity (if any) of the considered problem. It is noted that Problem $3$ is equivalent to Problem $i$ when $\lambda_i=1$ and $\lambda_j=0$,  $\forall i\ne j$. Here, we mean by equivalence that both problems have the same optimal solution.

\section{Solution of the Optimization Problem}
Problems 1, 2, and 3 are non-convex problems due to the non-convex constraints C1--C4. Besides, constraints C1, C3, and C4 involve infinitely many inequality constraints due to the continuity of the corresponding CSI uncertainty sets. To solve these problems efficiently, we first transform C1, C3, and C4 into equivalent linear matrix inequality (LMI) constraints. Then, Problems 1, 2, and 3 are solved by semidefinite programming (SDP) relaxation.

To facilitate the SDP relaxation, we define $\mathbf{W}_k=\mathbf{w}_k\mathbf{w}_k^H$ and rewrite Problems 1-3 in the following equivalent forms: \vspace*{-2mm}
\begin{eqnarray}
\label{eqn:prob1-rank-one}
&&\hspace*{-10mm}\text{\emph{Equivalent Problem 1:}} \notag \\
&&\hspace*{-1mm} \underset{\mathbf{W}_k,\mathbf{Z}\in\mathbb{H}^{N_\mathrm{T}},P_j}{\mino}\,\, \,\, \sum_{k=1}^{K}\Tr(\mathbf{W}_k)+\Tr(\mathbf{Z}) \notag \\[-1mm]
\notag\mbox{s.t.}
&&\hspace*{-5mm}\mbox{C1}\mbox{: }\underset{\Delta f_{j,k}\in{\Omega}_{j,k}}{\min} \frac{\Tr(\mathbf{H}_k\mathbf{W}_k)}{\overset{K}{\underset{r \neq k}{\sum}}\Tr(\mathbf{H}_k\mathbf{W}_r) + \overset{J}{\underset{j=1}{\sum}}P_j\abs{f_{j,k}}^2 + \Tr(\mathbf{H}_k\mathbf{Z})+\sigma_{\mathrm{n}_k}^2} \geq \Gamma^{\mathrm{DL}}_{\mathrm{req}_k},\,\, \forall k, j, \notag\\[-1mm]
&&\hspace*{-5mm}\mbox{C2}\mbox{: }\frac{P_j\Tr(\mathbf{G}_j\mathbf{V}_j)} {\overset{J}{\underset{n \neq j}{\sum}}P_n\abs{\mathbf{g}_n^H\mathbf{v}_j}^2\hspace*{-0.5mm}+\hspace*{-0.5mm}\Tr\hspace*{-1mm}\Big(\hspace*{-0.5mm}\rho\mathbf{V}_j\diag\Big(\mathbf{H}_{\mathrm{SI}}\mathbf{Z}\mathbf{H}_{\mathrm{SI}}^H \hspace*{-0.5mm}+\hspace*{-0.5mm}\overset{K}{\underset{k=1}{\sum}} \mathbf{H}_{\mathrm{SI}} \mathbf{W}_k\mathbf{H}_{\mathrm{SI}}^H\Big)\Big) \hspace*{-1mm}+\hspace*{-1mm}\sigma_{\mathrm{UL}}^2\norm{\mathbf{v}_j}^2} \hspace*{-1mm} \ge \hspace*{-1mm} \Gamma^{\mathrm{UL}}_{\mathrm{req}_j},\,\, \forall j, \notag\\
&&\hspace*{-5mm}\mbox{C3}\mbox{: }\underset{\Delta\mathbf{ L}_m\in \mathbf{\Omega}_{\mathrm{DL}_m}}{\max} \log_2\det(\mathbf{I}_{N_{\mathrm{R}}}+\mathbf{X}_m^{-1}\mathbf{L}_m^H\mathbf{W}_k\mathbf{L}_m) \le R_{\mathrm{tol}_{k,m}}^{\mathrm{DL}} ,\,\, \forall k, m,\notag\\[-0mm]
&&\hspace*{-5mm}\mbox{C4}\mbox{: }\underset{\Delta \mathbf{e}_{j,m} \in \mathbf{\Omega}_{\mathrm{UL}_{j,m}}}{\underset{\Delta\mathbf{ L}_m\in \mathbf{\Omega}_{\mathrm{DL}_m}}{\max}} \log_2\det(\mathbf{I}_{N_{\mathrm{R}}}+P_j\mathbf{X}_m^{-1}\mathbf{e}_{j,m}\mathbf{e}_{j,m}^H) \le R_{\mathrm{tol}_{j,m}}^{\mathrm{UL}} ,\,\, \forall j, m, \notag \\[-1mm]
&&\hspace*{-5mm}\mbox{C5}\mbox{: } P_j \geq 0,\,\, \forall j,\quad \mbox{C6}\mbox{: }\mathbf{Z} \succeq \mathbf{0},\quad
\mbox{C7}\mbox{: }\mathbf{W}_k \succeq \mathbf{0}, \forall k,\quad \mbox{C8}\mbox{: }\Rank(\mathbf{W}_k) \le 1, \forall k,
\end{eqnarray}
where $\mathbf{W}_k\succeq \mathbf{0}$, ${\mathbf{W}_k}\in \mathbb{H}^{N_\mathrm{T}}$, and $\Rank(\mathbf{W}_k) \le 1$ in (\ref{eqn:prob1-rank-one}) are imposed to guarantee that $\mathbf{W}_k=\mathbf{w}_k\mathbf{w}_k^H$ holds after optimization. \vspace*{-2mm}
\begin{eqnarray}
\label{eqn:UL-power-min-transformed}
&&\hspace*{-74mm}\text{\emph{Equivalent Problem 2:}} \notag \\[-3mm]
&&\hspace*{-15mm} \underset{\mathbf{W}_k,\mathbf{Z}\in\mathbb{H}^{N_\mathrm{T}},P_j}{\mino}\,\, \,\, \sum_{j=1}^{J}P_j \notag\\[-0.5mm]
\mbox{s.t.} &&\hspace*{-3mm}{\mbox{C1}}-{\mbox{C8}}.
\end{eqnarray}
\vspace*{-13mm}
\begin{eqnarray}
\label{eqn:MOP-transformed}
&&\hspace*{-56mm}\text{\emph{Equivalent Problem 3:}} \notag \\[-3mm]
&&\hspace*{8mm}\underset{\mathbf{W}_k,\mathbf{Z}\in\mathbb{H}^{N_\mathrm{T}},P_j,\tau}{\mino}\,\,\tau\nonumber\\[-0.5mm]
&&\hspace*{-6mm}\mbox{s.t.}\hspace*{15mm} {\mbox{C1}}-{\mbox{C8}},\notag\\[-4mm]
&&\hspace*{-6mm}{\mbox{C9}}\mbox{: }\lambda_i (Q_i-Q_i^*)\le \tau, \forall i\in\{1,2\},
\end{eqnarray}
where $\tau$ is an auxiliary optimization variable and (\ref{eqn:MOP-transformed}) is the epigraph representation of (\ref{eqn:MOP}).

Since Problem 3 is a generalization of Problems 1 and 2, we focus on solving Problem 3. Now, we introduce a Lemma which allows us to transform constraint ${\mbox{C1}}$ into an LMI.
\begin{Lem}[S-Procedure \cite{book:convex}] Let a function $f_m(\mathbf{x}),m\in\{1,2\},\mathbf{x}\in \mathbb{C}^{N\times 1},$ be defined as \vspace*{-2mm}
\begin{eqnarray}
\label{eqn:S-proc-function}f_m(\mathbf{x})=\mathbf{x}^H\mathbf{A}_m\mathbf{x}+2 \hspace*{0mm}\Re\hspace*{0mm} \{\mathbf{b}_m^H\mathbf{x}\}+c_m,
\end{eqnarray}
where $\mathbf{A}_m\in\mathbb{H}^N$, $\mathbf{b}_m\in\mathbb{C}^{N\times 1}$, and $c_m\in\mathbb{R}^{1\times 1}$. Then, the implication $f_1(\mathbf{x})\le 0\Rightarrow f_2(\mathbf{x})\le 0$  holds if and only if there exists a variable $\delta\ge 0$ such that
\begin{eqnarray}\label{eqn:S-proc-LMI}\delta
\begin{bmatrix}
       \mathbf{A}_1 & \mathbf{b}_1          \\
       \mathbf{b}_1^H & c_1           \\
           \end{bmatrix} -\begin{bmatrix}
       \mathbf{A}_2 & \mathbf{b}_2          \\
       \mathbf{b}_2^H & c_2           \\
           \end{bmatrix}          \succeq \mathbf{0}         ,
\end{eqnarray}
provided that there exists a point $\mathbf{\hat{x}}$ such that $f_k(\mathbf{\hat{x}})<0$.
\end{Lem}

To facilitate the presentation, we first define $\mathbf{f}_k\hspace*{-0.5mm}=\hspace*{-0.5mm}\big[f_{1,k},\ldots,f_{J,k}\big]^T$, $\mathbf{\hat f}_k\hspace*{-0.5mm}=\hspace*{-0.5mm}\big[\hat f_{1,k},\ldots,\hat f_{J,k}\big]^T$, $\Delta \mathbf{f}_k\hspace*{-0.5mm}=\hspace*{-0.5mm}\big[\Delta f_{1,k},\ldots,\Delta f_{J,k}\big]^T$, and $\mathbf{P}\hspace*{-0.5mm}=\hspace*{-0.5mm}\diag\big(P_1,\ldots,P_J\big)$,
where $\mathbf{f}_k$, $\mathbf{\hat f}_k$, and $\Delta \mathbf{f}_k$ denote the collections of the CCI channels, CCI channel estimates, and CCI estimation errors at DL user $k$, respectively. Hence, the collection of CCI channels at DL user $k$ can be modeled as \vspace*{-2mm}
\begin{eqnarray}
\label{eqn:CCI-colletion-vector}\mathbf{f}_{k}=\mathbf{\hat  f}_k  + \Delta \mathbf{f}_{k},\,   k\in\{1,\ldots,K\}, \,\,\,
{\Omega }_{k}\triangleq \Big\{ \mathbf{f}_{k}\in \mathbb{C}^{{J}\times 1}  :\norm{\Delta \mathbf{f}_{k}} \le \varepsilon_{k}\Big\},
\end{eqnarray}
where $\varepsilon_{k}^2=\sum_{j=1}^{J}\varepsilon_{j,k}^2$. As a result, by applying (\ref{eqn:CCI-colletion-vector}), ${\mbox{C1}}$ can be equivalently expressed as \vspace*{-2mm}
\begin{eqnarray}
\widetilde{\mbox{C1}}\mbox{: }0 \hspace*{-0.5mm} \ge\hspace*{-0.5mm} \Delta\mathbf{f}^H_{k}\mathbf{P}\Delta\mathbf{f}_{k}\hspace*{-0.5mm} + \hspace*{-0.5mm}2\Re\{\mathbf{\hat  f}^H_k\mathbf{P}\Delta\mathbf{f}_k\}
\hspace*{-0.5mm} +\hspace*{-0.5mm}\mathbf{\hat f}^H_k\mathbf{P}\mathbf{\hat f}_k\hspace*{-0.5mm}
+\hspace*{-0.5mm}\overset{K}{\underset{r \neq k}{\sum}}\Tr(\mathbf{H}_k\mathbf{W}_r)\hspace*{-0.5mm}
-\hspace*{-0.5mm}\frac{\Tr(\mathbf{H}_k\mathbf{W}_k)}{\Gamma_{\mathrm{req}_k}^{\mathrm{DL}}}\hspace*{-0.5mm}  +\hspace*{-0.5mm}\Tr(\mathbf{H}_k\mathbf{Z})\hspace*{-0.5mm} +\hspace*{-0.5mm}\sigma_{\mathrm{n}_k}^2\hspace*{-0.5mm}.\,\, \notag
\end{eqnarray}
By exploiting Lemma 1, we obtain the following implications: \vspace*{-2mm}
\begin{eqnarray}
\label{eqn:CCI-err-QC}\Delta \mathbf{f}_{k}^H\Delta \mathbf{f}_{k} - \varepsilon_{k}^2 \le 0
\Rightarrow   \widetilde{\mbox{C1}} \notag
\end{eqnarray}
holds if and only if there exists a variable $\delta_k \ge 0$ such that \vspace*{-2mm}
\begin{eqnarray}\label{eqn:LMI-C1}
\overline{\text{C1}}\mbox{: }\mathbf{R}_{\overline{\mathrm{C1}}_k}\big(\mathbf{W}_k,\mathbf{Z},P_j,\delta_k\big)\hspace*{106mm}\notag\\
=
          \begin{bmatrix}
       \delta_k\mathbf{I}_{J}\hspace*{-0.5mm}-\hspace*{-0.5mm}\mathbf{P} \hspace*{-1mm}&\hspace*{-1mm} -\mathbf{\hat f}_k          \\
       -\mathbf{\hat f}_k^H    \hspace*{-1mm} & \hspace*{-1mm} -\delta_k\varepsilon_k^2 \hspace*{-0.5mm}-\hspace*{-0.5mm}\sigma_{\mathrm{n}_k}^2 \hspace*{-0.5mm}-\hspace*{-0.5mm}  \mathbf{\hat f}_k^H \mathbf{P} \mathbf{\hat f}_k \hspace*{-0.5mm}-\hspace*{-0.5mm} \Tr(\mathbf{H}_k\mathbf{Z})        \\
           \end{bmatrix}
             \hspace*{-0.5mm}-\hspace*{-0.5mm} \mathbf{B}_{\mathbf{h}_k}^H\Big(\overset{K}{\underset{r \neq k}{\sum}}\mathbf{W}_r\hspace*{-0.5mm}-\hspace*{-0.5mm}\frac{\mathbf{W}_k}{\Gamma_{\mathrm{req}_k}^{\mathrm{DL}}}\Big)\mathbf{B}_{\mathbf{h}_k} \succeq \mathbf{0}, \forall k, j,
\end{eqnarray}
holds, where $\mathbf{B}_{\mathbf{h}_k}=\big[\mathbf{0}\quad\mathbf{h}_k\big]$.

Next, for handling the non-convex constraints ${\mbox{C3}}$ and ${\mbox{C4}}$, we establish the following proposition for facilitating the constraint transformation for the considered optimization problem.

\begin{Prop}
For $R_{\mathrm{tol}_{k,m}}^{\mathrm{DL}} > 0$ and $R_{\mathrm{tol}_{j,m}}^{\mathrm{UL}} > 0$, we have the following implications for constraints ${\mbox{C3}}$ and ${\mbox{C4}}$ of equivalent Problems 1-3, respectively:
\begin{eqnarray}
\label{eqn:DL-tol-rate-QC}{\mbox{C3}}&\Rightarrow&{\widetilde{\mbox{C3}}}\mbox{: }\mathbf{L}_m^H\mathbf{W}_k\mathbf{L}_m \preceq \xi_{k,m}^{\mathrm{DL}}\mathbf{X}_m,\,\, \forall \mathbf{L}_m\in \mathbf{\Omega}_{\mathrm{DL}_m},\, \forall k, m,\\
\label{eqn:UL-tol-rate-QC}{\mbox{C4}}&\Leftrightarrow&{\widetilde{\mbox{C4}}}\mbox{: }P_j\mathbf{e}_{j,m}\mathbf{e}_{j,m}^H \preceq \xi_{j,m}^{\mathrm{UL}} \mathbf{X}_m ,\,\, \forall \mathbf{e}_{j,m} \in \mathbf{\Omega}_{\mathrm{UL}_{j,m}},\, \forall \mathbf{L}_m\in \mathbf{\Omega}_{\mathrm{DL}_m},\, \forall j, m,
\end{eqnarray}
where $\xi_{k,m}^{\mathrm{DL}}=2^{R_{\mathrm{tol}_{k,m}}^{\mathrm{DL}}}-1$ and $\xi_{j,m}^{\mathrm{UL}}=2^{R_{\mathrm{tol}_{j,m}}^{\mathrm{UL}}}-1$. We note that ${\mbox{C3}}$ and ${\widetilde{\mbox{C3}}}$ are equivalent respectively if $\Rank(\mathbf{W}_k) \le 1$. Besides, ${\mbox{C4}}$ and ${\widetilde{\mbox{C4}}}$ are always equivalent.
\end{Prop}
\emph{\quad Proof: } Please refer to Appendix A. \hfill\qed

Although ${\widetilde{\mbox{C3}}}$ and ${\widetilde{\mbox{C4}}}$ are convex LMI constraints which are less difficult to handle compared to ${\mbox{C3}}$ and ${\mbox{C4}}$, they still involve an infinite number of inequality constraints. To circumvent this difficulty, we introduce the following Lemma to further simplify ${\widetilde{\mbox{C3}}}$ and ${\widetilde{\mbox{C4}}}$.

\begin{Lem}[Generalized S-Procedure \cite{luo2004multivariate}] \label{lem:gernal-s-proc}Let $f(\mathbf{X})=\mathbf{X}^H\mathbf{A}\mathbf{X}+\mathbf{X}^H\mathbf{B}+\mathbf{B}^H\mathbf{X}+\mathbf{C}$, and $\mathbf{D}\succeq \mathbf{0}$. For some $t \ge 0$, $f(\mathbf{X})\succeq \mathbf{0}, \forall \mathbf{X} \in \Big\{\mathbf{X} |\Tr(\mathbf{DXX}^H) \le 1 \Big\}$, is equivalent to
\begin{eqnarray}
\label{eqn:General-S-proc-LMI}
&&\begin{bmatrix}
       \mathbf{C} & \mathbf{B}^H          \\
       \mathbf{B} & \mathbf{A}           \\
           \end{bmatrix} -t\begin{bmatrix}
       \mathbf{I} & \mathbf{0}          \\
       \mathbf{0} & -\mathbf{D}           \\
           \end{bmatrix}          \succeq \mathbf{0}.
\end{eqnarray}
\end{Lem}

As a result, we first substitute $\mathbf{L}_m\hspace*{-0mm}=\hspace*{-0mm}\mathbf{\hat L}_m \hspace*{-0mm}+ \hspace*{-0mm}\Delta\mathbf{L}_m$ into (\ref{eqn:DL-tol-rate-QC}) and express constraint $\widetilde{\mbox{C3}}$ as \vspace*{-2mm}
\begin{eqnarray}
\label{eqn:BS-EVE-err-QC}\mathbf{0} &\preceq& \Delta\mathbf{L}_m^H(\xi_{k,m}^{\mathrm{DL}}\mathbf{Z}-\mathbf{W}_k)\Delta\mathbf{L}_m +\Delta\mathbf{L}_m^H(\xi_{k,m}^{\mathrm{DL}}\mathbf{Z}-\mathbf{W}_k)\mathbf{\hat L}_m \notag\\
&+&\mathbf{\hat L}_m^H(\xi_{k,m}^{\mathrm{DL}}\mathbf{Z}-\mathbf{W}_k)\Delta\mathbf{L}_m +\mathbf{\hat L}_m^H(\xi_{k,m}^{\mathrm{DL}}\mathbf{Z}-\mathbf{W}_k)\mathbf{\hat L}_m +\xi_{k,m}^{\mathrm{DL}}\sigma_{\mathrm{E}_m}^2\mathbf{I}_{N_\mathrm{R}},\forall k, m,
\end{eqnarray}
for $\Delta\mathbf{L}_m \in \Big\{\Delta\mathbf{L}_m |\Tr(\varepsilon_{\mathrm{DL}_m}^{-2}\Delta\mathbf{L}_m\Delta\mathbf{L}^H_m) \le 1 \Big\}$.
Then, by applying Lemma 2, constraint $\widetilde{\mbox{C3}}$ is equivalently represented as \vspace*{-2mm}
\begin{eqnarray}\label{eqn:LMI-C3}
\hspace*{-50mm}\overline{\text{C3}}\mbox{: } \mathbf{R}_{{\overline{\mathrm{C3}}}_{k,m}}\big(\mathbf{W}_k,\mathbf{Z},t_{k,m}\big)\hspace*{108mm}\notag\\
\hspace*{-1mm}=\hspace*{-1mm}
          \begin{bmatrix}
       \xi_{k,m}^{\mathrm{DL}}\mathbf{\hat L}_m^H\mathbf{Z}\mathbf{\hat L}_m\hspace*{-0.5mm}+\hspace*{-0.5mm}(\xi_{k,m}^{\mathrm{DL}}\sigma_{\mathrm{E}_m}^2-t_{k,m})\mathbf{I}_{N_\mathrm{R}} \hspace*{-1mm}&\hspace*{-1mm} \xi_{k,m}^{\mathrm{DL}}\mathbf{\hat L}_m^H\mathbf{Z}          \\
       \xi_{k,m}^{\mathrm{DL}}\mathbf{Z}\mathbf{\hat L}_m    \hspace*{-1mm}&\hspace*{-1mm}   \xi_{k,m}^{\mathrm{DL}}\mathbf{Z}\hspace*{-0.5mm}+\hspace*{-0.5mm}  t_{k,m}\varepsilon_{\mathrm{DL}_m}^{-2}\mathbf{I}_{N_\mathrm{T}}     \\
           \end{bmatrix}
            \hspace*{-1mm}-\hspace*{-1mm}\mathbf{B}_{\mathbf{L}_m}^H\mathbf{W}_k\mathbf{B}_{\mathbf{L}_m}\succeq \mathbf{0}, \forall k, m,
\end{eqnarray}
for $t_{k,m} \ge 0$, $\forall k, m$, where $\mathbf{B}_{\mathbf{L}_m}=\big[\mathbf{\hat L}_m\quad\mathbf{I}_{N_\mathrm{T}}\big]$.
On the other hand, for constraint ${\widetilde{\mbox{C4}}}$, two estimation error variables are involved, namely $\Delta\mathbf{e}_{j,m}$ and $\Delta\mathbf{L}_m$, and Lemma 2 cannot be directly applied. Hence, we introduce a slack matrix variable $\mathbf{M}_{j,m} \in \mathbb{H}^{N_{\mathrm{R}}}$ to handle the coupled estimation error variables in constraint ${\widetilde{\mbox{C4}}}$. In particular, constraint ${\widetilde{\mbox{C4}}}$ can be equivalently represented by \vspace*{-2mm}
\begin{eqnarray}
\label{eqn:decoupled-UL-QC1}&&{\widetilde{\mbox{C4a}}}\mbox{: }P_j\mathbf{e}_{j,m}\mathbf{e}_{j,m}^H \preceq \mathbf{M}_{j,m},\,\, \forall \mathbf{e}_{j,m} \in \mathbf{\Omega}_{\mathrm{UL}_{j,m}},\, \forall j, m,\\
\label{eqn:decoupled-UL-QC2}&&{\widetilde{\mbox{C4b}}}\mbox{: }\mathbf{M}_{j,m} \preceq (\xi_{j,m}^{\mathrm{UL}}-1) \mathbf{X}_m,\,\, \forall \mathbf{L}_m\in \mathbf{\Omega}_{\mathrm{DL}_m},\, \forall j, m.
\end{eqnarray}

\begin{Prop}
Constraint ${\widetilde{\mbox{C4}}}$ holds if there exists a Hermitian matrix $\mathbf{M}_{j,m}\in \mathbb{H}^{N_{\mathrm{R}}}$, $j\in\{1,\ldots,J\}$, $m\in\{1,\ldots,M\}$ which meets constraints ${\widetilde{\mbox{C4a}}}$ and ${\widetilde{\mbox{C4b}}}$.
\end{Prop}

\emph{\quad Proof: } Please refer to Appendix B. \hfill\qed

Then, we apply Lemma 2 to constraints ${\widetilde{\mbox{C4a}}}$ and ${\widetilde{\mbox{C4b}}}$ in a similar manner as for handling ${\widetilde{\mbox{C3}}}$ and obtain the following equivalent LMIs for ${\widetilde{\mbox{C4a}}}$ and ${\widetilde{\mbox{C4b}}}$, respectively: \vspace*{-2mm}
\begin{eqnarray}\label{eqn:LMI-C6}
\overline{\text{C4}}\mbox{a: }\mathbf{R}_{{\overline{\mathrm{C4}}\mathrm{a}}_{j,m}}\hspace*{-0.5mm}\big(\hspace*{-0.5mm} \mathbf{M}_{j,m},P_j,\alpha_{j,m}\hspace*{-0.5mm} \big)\hspace*{-0.8mm} =\hspace*{-1.5mm}
          \begin{bmatrix}
       -P_j\mathbf{\hat e}_{j,m}\mathbf{\hat e}_{j,m}^H \hspace*{-1.5mm}+\hspace*{-1mm} \mathbf{M}_{j,m}\hspace*{-1.5mm} -\hspace*{-0.5mm} \alpha_{j,m}\mathbf{I}_{N_\mathrm{R}}  \hspace*{-1mm}  & \hspace*{-1mm} -P_j\mathbf{\hat e}_{j,m}          \\
       \hspace*{-1.5mm} -P_j\mathbf{\hat e}_{j,m}^H  \hspace*{-1mm}  & \hspace*{-3.5mm} -P_j\hspace*{-0.5mm}+\hspace*{-0.5mm} \alpha_{j,m}\varepsilon_{\mathrm{UL}_{j,m}}^{-2}      \\
           \end{bmatrix}\hspace*{-2mm}\succeq\hspace*{-1mm} \mathbf{0}, \forall j, m,
\end{eqnarray}
for $\alpha_{j,m} \ge 0$, $\forall j, m$,
and \vspace*{-2mm}
\begin{eqnarray}\label{eqn:LMI-C7}
&&\overline{\text{C4}}\mbox{b: }\mathbf{R}_{{\overline{\mathrm{C4}}\mathrm{b}}_{j,m}}\hspace*{-0.5mm}\big(\mathbf{Z},\mathbf{M}_{j,m},\beta_{j,m}\big)\notag\\
=&&\hspace*{-4mm}
          \begin{bmatrix}
       \xi_{j,m}^{\mathrm{UL}}\mathbf{\hat L}_m^H\mathbf{Z}\mathbf{\hat L}_m \hspace*{-0.5mm}+\hspace*{-0.5mm}(\xi_{j,m}^{\mathrm{UL}}\sigma_{\mathrm{E}_m}^2 \hspace*{-0.5mm}- \hspace*{-0.5mm} \beta_{j,m})\mathbf{I}_{N_\mathrm{R}}\hspace*{-0.5mm} -\hspace*{-0.5mm} \mathbf{M}_{j,m} \hspace*{-1mm} & \hspace*{-1mm} \xi_{j,m}^{\mathrm{UL}}\mathbf{\hat L}_m^H\mathbf{Z}          \\
       \xi_{j,m}^{\mathrm{UL}}\mathbf{Z}\mathbf{\hat L}_m   \hspace*{-1mm} & \hspace*{-1mm}  \xi_{j,m}^{\mathrm{UL}}\mathbf{Z} \hspace*{-0.5mm}+\hspace*{-0.5mm}  \beta_{j,m}\varepsilon_{\mathrm{DL}_m}^{-2}\mathbf{I}_{N_\mathrm{T}}     \\
           \end{bmatrix}  \succeq \mathbf{0}, \forall j, m,
\end{eqnarray}
for $\beta_{j,m} \ge 0$, $\forall j, m$.

The remaining non-convex constraint in (\ref{eqn:MOP-transformed}) is the rank-one constraint C8. Solving such a rank-constrained problem is known to be NP-hard \cite{gartner2012approximation}. Hence, to obtain a tractable solution, we relax constraint C8: $\Rank(\mathbf{W}_k) \le 1$ by removing it from the problem formulation, such that the considered problem becomes a convex SDP and is given by \vspace*{-10mm}

\begin{eqnarray}
\label{eqn:SDR-MOP-robust}&&\hspace*{-53mm}\underset{P_j,\tau,\delta_k, t_{k,m},\alpha_{j,m},\beta_{j,m}} {\underset{\mathbf{W}_k,\mathbf{Z}\in\mathbb{H}^{N_\mathrm{T}},\mathbf{M}_{j,m}\in\mathbb{H}^{N_\mathrm{R}},}{\mino}}\,\, \tau \notag \\[2mm]
\notag \mbox{s.t.} \hspace*{5mm}{\mbox{C2}},{\mbox{C5}},{\mbox{C6}},{\mbox{C7}},{\mbox{C9}}, \hspace*{57mm}
&& \hspace*{-30mm}\mbox{C10: }\delta_k,t_{k,m},\alpha_{j,m}, \beta_{j,m} \ge 0,\forall k,j,m,\\
\overline{\text{C1}}\mbox{: }\mathbf{R}_{\overline{\mathrm{C1}}_k}\big(\mathbf{W}_k,\mathbf{Z},P_j,\delta_k\big)\succeq \mathbf{0}, \forall k,\notag\ \hspace*{39mm}
&& \hspace*{-30mm}\overline{\text{C3}}\mbox{: }\mathbf{R}_{{\overline{\mathrm{C3}}}_{k,m}}\big(\mathbf{W}_k,\mathbf{Z},t_{k,m}\big)\succeq \mathbf{0},\forall k,m, \notag\\
\overline{\text{C4}}\mbox{a: }\mathbf{R}_{{\overline{\mathrm{C4}}\mathrm{a}}_{j,m}}\big(\mathbf{M}_{j,m},P_j,\alpha_{j,m}\big)\succeq \mathbf{0},\forall j,m,  \hspace*{28mm}
&& \hspace*{-30mm}\overline{\text{C4}}\mbox{b: }\mathbf{R}_{{\overline{\mathrm{C4}}\mathrm{b}}_{j,m}}\big(\mathbf{Z},\mathbf{M}_{j,m},\beta_{j,m}\big)\succeq \mathbf{0},\forall j, m.
\end{eqnarray}
The relaxed convex problem in (\ref{eqn:SDR-MOP-robust}) can be solved efficiently by standard convex program solvers such as CVX \cite{website:CVX}. Besides, if the solution obtained for a relaxed SDP problem is a rank-one matrix, i.e., $\Rank(\mathbf{W}_k) = 1$ for $\mathbf{W}_k\ne\mathbf{0}, \, \forall k$, then it is also the optimal solution of the original problem. Next, we verify the tightness of the adopted SDP relaxation in the following theorem.
\begin{Thm}\label{thm:rankone_condition}
Assuming the considered problem is feasible, for $\Gamma^{\mathrm{DL}}_{\mathrm{req}_k} > 0$, we can always obtain or construct a rank-one optimal matrix $\mathbf{W}_k^*$ which is an optimal solution for (\ref{eqn:SDR-MOP-robust}).
\end{Thm}
\emph{\quad Proof: } Please refer to Appendix C. \hfill\qed

By Theorem 1, the optimal beamforming vector $\mathbf{w}_k^*$ can be recovered from $\mathbf{W}_k^*$ by performing eigenvalue decomposition of $\mathbf{W}_k^*$ and selection of the principle eigenvector as $\mathbf{w}_k^*$.

\vspace*{-2mm}
\section{Results}
In this section, we investigate the performance of the proposed multi-objective optimization based resource allocation scheme through simulations. The most important simulation parameters are specified in Table \ref{tab:parameters}. There are $K=3$ DL users, $J=7$ UL users, and $M=2$ potential eavesdroppers in a cell. The users and potential eavesdroppers are randomly and uniformly distributed between the reference distance of $30$ meters and the maximum service distance of $600$ meters. The FD BS is located at the center of the cell and equipped with $N_\mathrm{T}$ antennas. Each potential eavesdropper is equipped with $N_\mathrm{R}=2$ antennas. The small scale fading of the DL channels, UL channels, CCI channels, and eavesdropping channels are modeled as independent and identically distributed Rayleigh fading. The multipath fading coefficients of the SI channel are generated as independent and identically distributed Rician random variables with Rician factor $5$ dB. To facilitate the presentation, we define the maximum normalized estimation error of the CCI channels, DL eavesdropping channels, and UL eavesdropping channels as $\frac{\varepsilon_{j,k}^2}{\abs{f_{j,k}}^2}=\kappa_{j,k}^2$,
$\frac{\varepsilon_{\mathrm{DL}_{m}}^2}{\norm{\mathbf{L}_{m}}^2_{F}}=\kappa_{\mathrm{DL}_{m}}^2$, and
$\frac{\varepsilon_{\mathrm{UL}_{j,m}}^2}{\norm{\mathbf{e}_{j,m}}^2}=\kappa_{\mathrm{UL}_{j,m}}^2$, respectively. Furthermore, we assume that different channels have the same maximum normalized estimation error, i.e., $\kappa_{j,k}^2=\kappa_{\mathrm{DL}_{m}}^2=\kappa_{\mathrm{UL}_{j,m}}^2=\kappa_{\mathrm{est}}^2$. In addition, we assume that all DL users and all UL users require the same minimum SINRs, respectively, i.e., $\Gamma^{\mathrm{DL}}_{\mathrm{req}_k}=\Gamma^{\mathrm{DL}}_{\mathrm{req}}$ and $\Gamma^{\mathrm{UL}}_{\mathrm{req}_j}=\Gamma^{\mathrm{UL}}_{\mathrm{req}}$.

\begin{table}[t]\vspace*{-7mm}\caption{System parameters.}\vspace*{-2mm}\label{tab:parameters} 
\newcommand{\tabincell}[2]{\begin{tabular}{@{}#1@{}}#2\end{tabular}}
\centering
\begin{tabular}{|l|l|}\hline
\hspace*{-1mm}Carrier center frequency and system bandwidth & $1.9$ GHz and  $200$ kHz \\
\hline
\hspace*{-1mm}Path loss exponent and SI cancellation constant, $\rho$ &  \mbox{$3.6$} and  \mbox{$-80$ dB} \cite{FDRad}  \\
\hline
\hspace*{-1mm}DL user noise power and UL BS noise power, $\sigma_{\mathrm{n}_k}^2$ and $\sigma_{\mathrm{UL}}^2$ &  \mbox{$-100$ dBm} and \mbox{$-110$ dBm}\footnotemark{}  \\
\hline
\hspace*{-1mm}Potential eavesdropper noise power, $\sigma_{\mathrm{E}_m}^2$, and BS antenna gain &  \mbox{$-100$ dBm} and  \mbox{$10$ dBi}   \\
\hline
\hspace*{-1mm}Maximum tolerable data rate at potential eavesdroppers for DL users, $R_{\mathrm{tol}_{k,m}}^{\mathrm{DL}}$&  \mbox{$1$ bit/s/Hz}  \\
\hline
\hspace*{-1mm}Maximum tolerable data rate at potential eavesdroppers for UL users, $R_{\mathrm{tol}_{j,m}}^{\mathrm{UL}}$&  \mbox{$1$ bit/s/Hz}  \\
\hline
\end{tabular}
\vspace*{-6mm}
\end{table}
\vspace*{-1mm}
\footnotetext{The DL user and potential eavesdropper noise powers are higher than the UL noise power at the BS since the BS is usually equipped with a high quality LNA \cite{holma2009lte}.}

\vspace*{-4mm}
\subsection{Transmit Power Trade-off Region}
\begin{figure}[t]
\centering \vspace*{-10mm}
\includegraphics[width=4.2in]{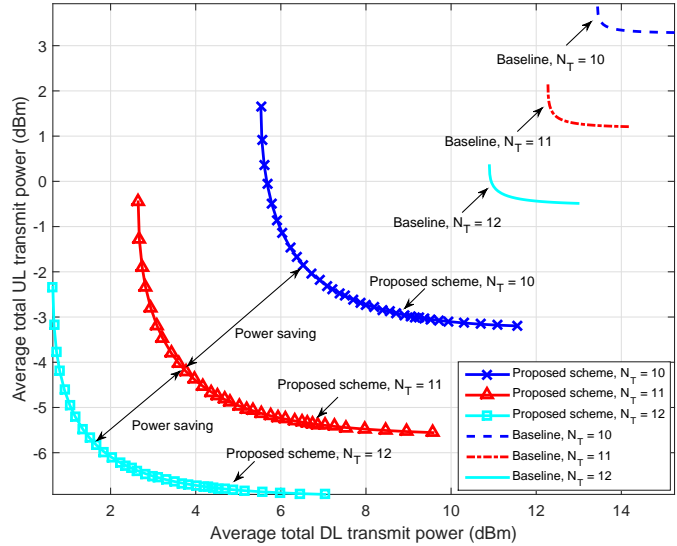}\vspace*{-5mm}
\caption{Average system objective trade-off region achieved by the proposed resource allocation scheme. The double-sided arrows indicate the power saving due to additional antennas.} \label{fig:tradeoff-err5}\vspace*{-8mm}
\end{figure}

In Figure \ref{fig:tradeoff-err5}, we study the trade-off between the DL and the UL total transmit powers for a maximum normalized channel estimation error of $\kappa_{\mathrm{est}}^2=5\%$ and different numbers of antennas at the FD BS. The trade-off region is obtained by solving (\ref{eqn:SDR-MOP-robust}) for different values of $0\leq \lambda_i \leq 1, \forall i \in \{1,2\}$, i.e., the $\lambda_i$ are varied uniformly using a step size of $0.01$ subject to $\sum_i \lambda_i=1$. We assume a minimum required DL SINR of $\Gamma^{\mathrm{DL}}_{\mathrm{req}}=10$ dB and a minimum required UL SINR of $\Gamma^{\mathrm{UL}}_{\mathrm{req}}=5$ dB. It can be observed from Figure \ref{fig:tradeoff-err5} that the total UL transmit power is a monotonically decreasing function with respect to the total DL transmit power. In other words, minimizing the total UL power consumption leads to a higher power consumption in the DL and vice versa. This result confirms that the minimization of the total UL transmit power and the total DL transmit power are conflicting design objectives. For the case of $N_\mathrm{T}=12$, $5$ dB in UL transmit power can be saved by increasing the total DL transmit power by $6$ dB. In addition, Figure 2 also indicates that a significant amount of transmit power can be saved in the FD system by increasing the number of BS antennas. This is due to the fact that the extra degrees freedom offered by the additional antennas facilitate a more power efficient resource allocation. However, due to channel hardening, there is a diminishing return in the power saving as the number of antennas at the FD BS increases \cite{book:david_wirelss_com}.

For comparison, we consider a baseline scheme. For the baseline scheme, we adopt ZF-BF as DL transmission scheme such that the multiuser interference is avoided at the DL legitimate users. In particular, the direction of beamformer $\mathbf{w}_k$ for legitimate user $k$ is fixed and lays in the null space of the other DL legitimate users' channels. Then, we jointly optimize $\mathbf{Z}$, $P_j$, and the power allocated to $\mathbf{w}_k$ under the MOOP formulation subject to the same constraints as in (\ref{eqn:SDR-MOP-robust}) via SDP relaxation.
Figure \ref{fig:tradeoff-err5} depicts the trade-off region for the baseline resource allocation scheme for comparison.
As can be seen, the trade-off regions achieved by the baseline scheme are above the curves for the proposed optimal scheme. This indicates that the proposed resource allocation scheme is more power efficient than the baseline scheme for both DL and UL transmission. Indeed, the proposed resource allocation scheme fully exploits the available degrees of freedom to perform globally optimal resource allocation. On the contrary, for the baseline scheme, the transmitter is incapable of fully utilizing the available degrees of freedom since the direction of the transmit beamformer $\mathbf{w}_k$ is fixed. Specifically, the fixed beamformer $\mathbf{w}_k$ can cause severe SI and increase the risk of eavesdropping which results in a high power consumption for UL transmission and the AN.
Besides, the trade-off region for the baseline scheme is strictly smaller than that of the proposed optimal scheme. For instance, when $N_\mathrm{T}=12$, the baseline scheme can save only $1$ dB of UL transmit power by increasing the total DL transmit power by $2$ dB, due to the limited flexibility of the baseline scheme in handling the interference.

We note that we also considered two other baseline schemes. For the first scheme, we adopted an isotropic radiation pattern for $\mathbf{Z}$ and optimized $\mathbf{W}_k$ and $P_j$. For the second scheme, we considered a wireless communication system with an HD BS. However, both of these schemes could not satisfy the QoS requirements in constraints C1 and C2 for the adopted simulation setting. Therefore, performance results for these two schemes are not shown in the paper.

\vspace*{-4mm}
\subsection{Average Total Transmit Power versus Minimum Required SINR}
\begin{figure}[t]
 \centering\vspace*{-6mm}
 \begin{minipage}[b]{0.45\linewidth} \hspace*{-1cm}
\includegraphics[width=3.7 in]{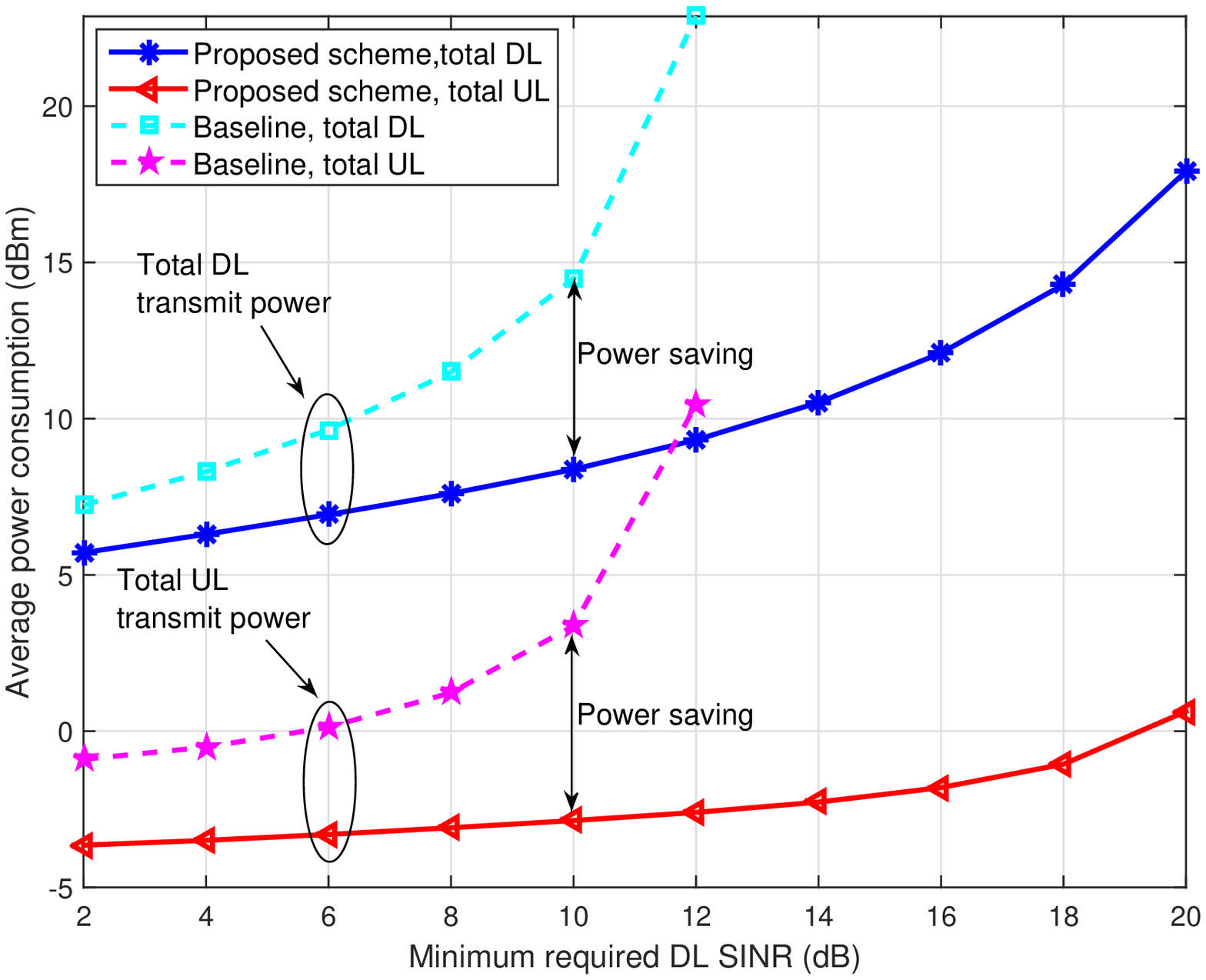}\vspace*{-8mm}
\caption{Average power consumption (dBm) versus the minimum required DL SINR (dB), $\Gamma^{\mathrm{DL}}_{\mathrm{req}}$, for different resource allocation schemes. } \label{fig:fig_total_power_vs_DL_sinr}
 \end{minipage}\hspace*{1.1cm}
 \begin{minipage}[b]{0.45\linewidth} \hspace*{-1cm}
\includegraphics[width=3.7 in]{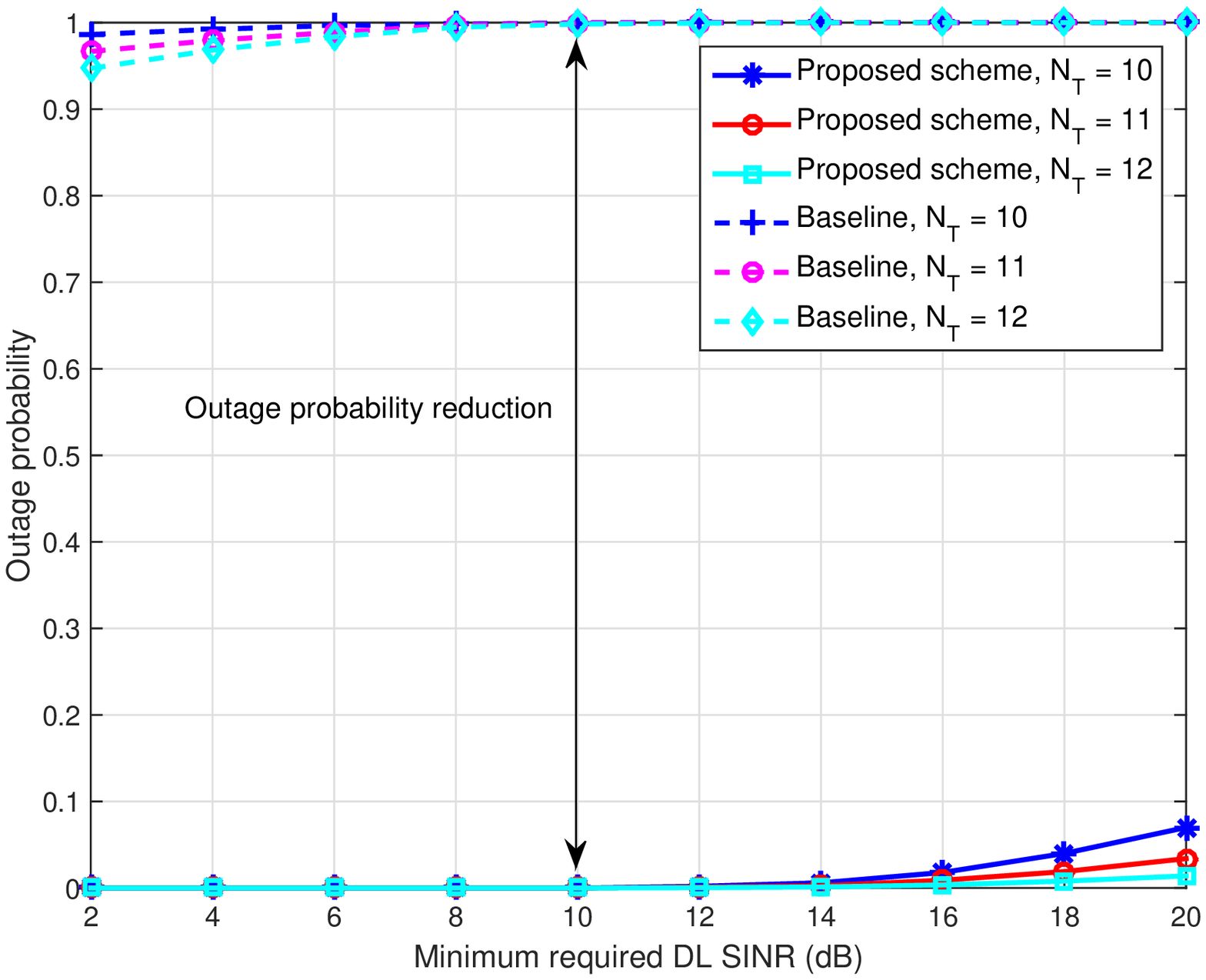}\vspace*{-8mm}
\caption{Outage probability versus the minimum required DL SINR (dB), $\Gamma^{\mathrm{DL}}_{\mathrm{req}}$, for different resource allocation schemes.} \label{fig:fig_outage_vs_probability}
 \end{minipage}\vspace*{-8mm}
\end{figure}

In Figure \ref{fig:fig_total_power_vs_DL_sinr}, we investigate the average power consumption of the DL and UL users versus the minimum required DL SINR, $\Gamma^{\mathrm{DL}}_{\mathrm{req}}$, for a minimum required UL SINR of $\Gamma^{\mathrm{UL}}_{\mathrm{req}}=5$ dB and a maximum normalized channel estimation error of $\kappa_{\mathrm{est}}^2=5\%$. The FD BS is equipped with $N_\mathrm{T}=10$ antennas. We select the resource allocation policy with $\lambda_1=0.1$ and $\lambda_2=0.9$ which indicates that the system operator gives priority to total UL transmit power minimization. It can be observed that both the DL and the UL power consumption increase with $\Gamma^{\mathrm{DL}}_{\mathrm{req}}$. However, the DL power consumption grows more rapidly than the UL power consumption. The reason behind this is threefold. First, a higher DL transmit power is required to fulfill the more stringent DL QoS requirements.
Second, a higher AN transmit power is required to neutralize the higher potential for information leakage. In particular, with the growth of the DL transmit power, the channel capacity between the FD BS and the potential eavesdroppers increases. Hence, the FD BS also has to allocate more power to the AN to prevent interception by potential eavesdroppers which results in a higher DL power consumption.
Third, the SI becomes more severe for higher DL transmit powers. Specifically, since the FD BS tries to keep the total UL transmit power low to avoid strong CCI for increasing required SINR, more degrees of freedom at the FD BS have to be utilized to suppress the SI to improve the receive SINR of the UL users. As a result, fewer degrees of freedom are available for reducing the DL transmit power consumption causing even higher DL transmit powers. On the other hand, the proposed scheme provides substantial power savings compared to the baseline scheme in both DL and UL for the entire range of $\Gamma^{\mathrm{DL}}_{\mathrm{req}}$. Also, the baseline scheme cannot satisfy the QoS requirements when $\Gamma^{\mathrm{DL}}_{\mathrm{req}}$ is larger than $12$ dB. In Figure \ref{fig:fig_outage_vs_probability}, we study the outage probability for the proposed scheme and the baseline scheme versus the minimum required DL SINR for a maximum normalized channel estimation error of $\kappa_{\mathrm{est}}^2=5\%$, a resource allocation policy with $\lambda_1=0.1$ and $\lambda_2=0.9$, and different numbers of antennas. An outage event occurs whenever the problem in \eqref{eqn:SDR-MOP-robust} is infeasible. As can be observed, a large outage probability reduction can be achieved with the proposed scheme compared to the baseline scheme. For the case of $\Gamma^{\mathrm{DL}}_{\mathrm{req}}=12$ dB and $N_\mathrm{T}=10$, the outage probability for the proposed scheme is only $0.5\%$, whereas the outage probability for the baseline scheme is $99.3\%$. These results confirm that the proposed scheme is more robust and reliable in the presence of imperfect CSI compared to the baseline scheme.

\begin{figure}[t]
 \centering \vspace*{-10mm}
\includegraphics[width=4.2in]{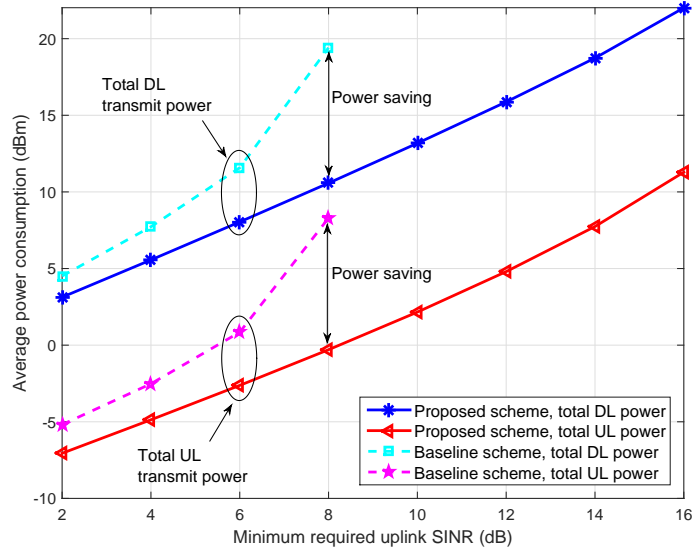} \vspace*{-5mm}
\caption{Average power consumption (dBm) versus the minimum required UL SINR (dB), $\Gamma^{\mathrm{UL}}_{\mathrm{req}}$, for different resource allocation schemes.} \label{fig:fig_total_power_vs_UL_sinr}\vspace*{-8mm}
\end{figure}

In Figure \ref{fig:fig_total_power_vs_UL_sinr}, we study the power consumption of the DL and UL users versus the minimum required UL SINR, $\Gamma^{\mathrm{UL}}_{\mathrm{req}}$, for a minimum required DL SINR of $\Gamma^{\mathrm{DL}}_{\mathrm{req}}=10$ dB and a maximum normalized channel estimation error of $\kappa_{\mathrm{est}}^2=5\%$. The FD BS is equipped with $N_\mathrm{T}=10$ antennas. The other system parameters are identical to those adopted in Figure \ref{fig:fig_total_power_vs_DL_sinr}. As can be observed, the total transmit power of DL and UL increase as the minimum required UL SINR increases. In fact, for the UL users, they need to transmit with higher power to fulfill the more stringent UL QoS requirements. Furthermore, increasing the total UL transmit power causes more CCI to the DL users. Hence, the FD BS has to allocate more power to the DL information signals to satisfy the DL QoS requirements. Moreover, the higher UL and DL information signal powers increase the susceptibility to eavesdropping for both UL and DL users. Hence, the FD BS also has to increase the AN power to guarantee DL and UL transmission security.
On the other hand, Figure \ref{fig:fig_total_power_vs_UL_sinr} clearly shows that the proposed scheme provides significant power savings compared to the baseline scheme in all considered scenarios. In fact, the baseline scheme cannot find a feasible solution for $\Gamma^{\mathrm{UL}}_{\mathrm{req}} > 8$ dB.

\vspace*{-4mm}
\subsection{Average User Secrecy Rate versus Minimum Required SINR}
\begin{figure}[t]
 \centering\vspace*{-6mm}
 \begin{minipage}[b]{0.45\linewidth} \hspace*{-1cm}
\includegraphics[width=3.7 in]{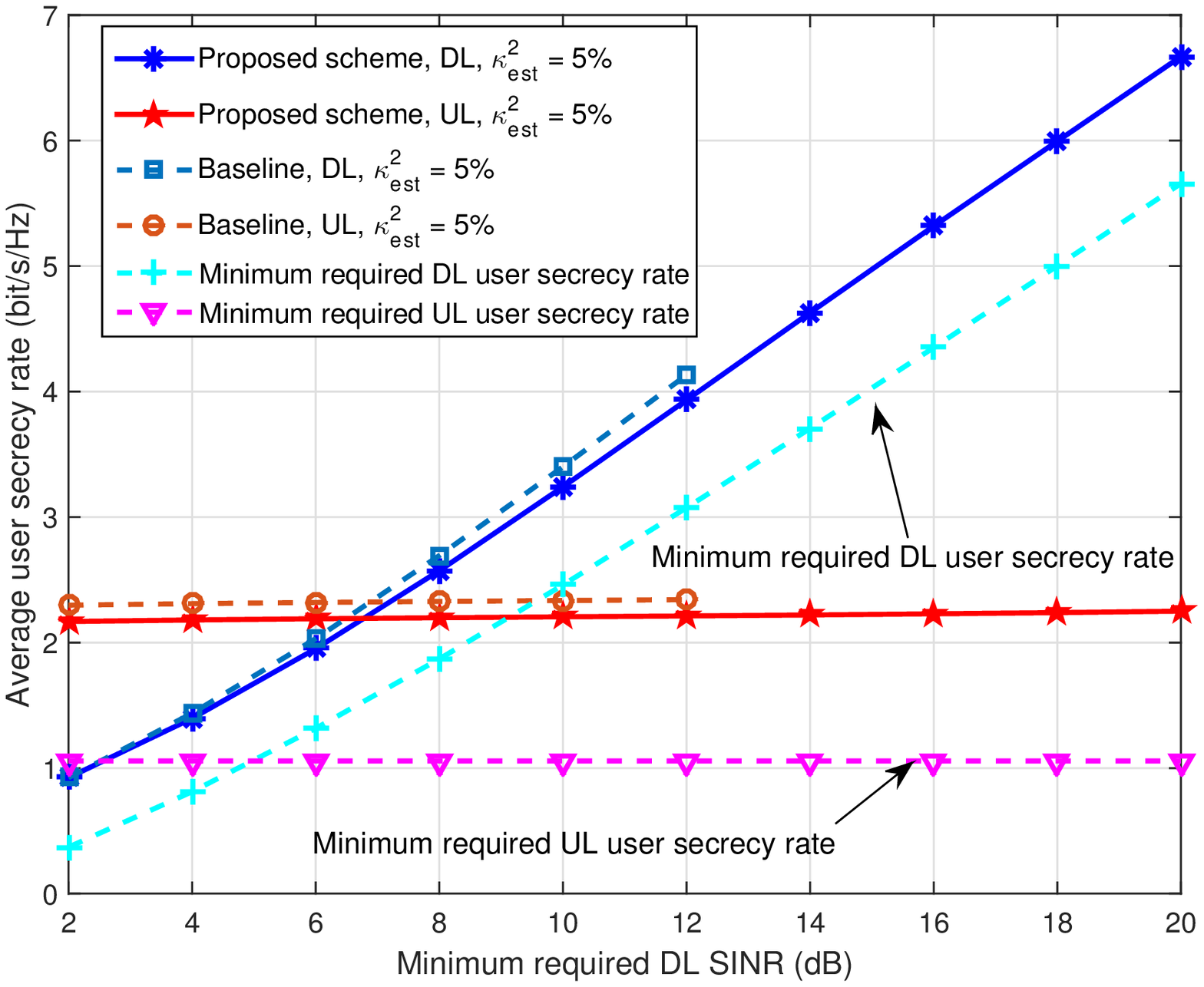}\vspace*{-8mm}
\caption{Average user secrecy rate (bits/s/Hz) versus the minimum required DL SINR (dB), $\Gamma^{\mathrm{DL}}_{\mathrm{req}}$.} \label{fig:SecrecyCap_vs_DL_SINR_err5_10}
 \end{minipage}\hspace*{1.1cm}
 \begin{minipage}[b]{0.45\linewidth} \hspace*{-1cm}
\includegraphics[width=3.7 in]{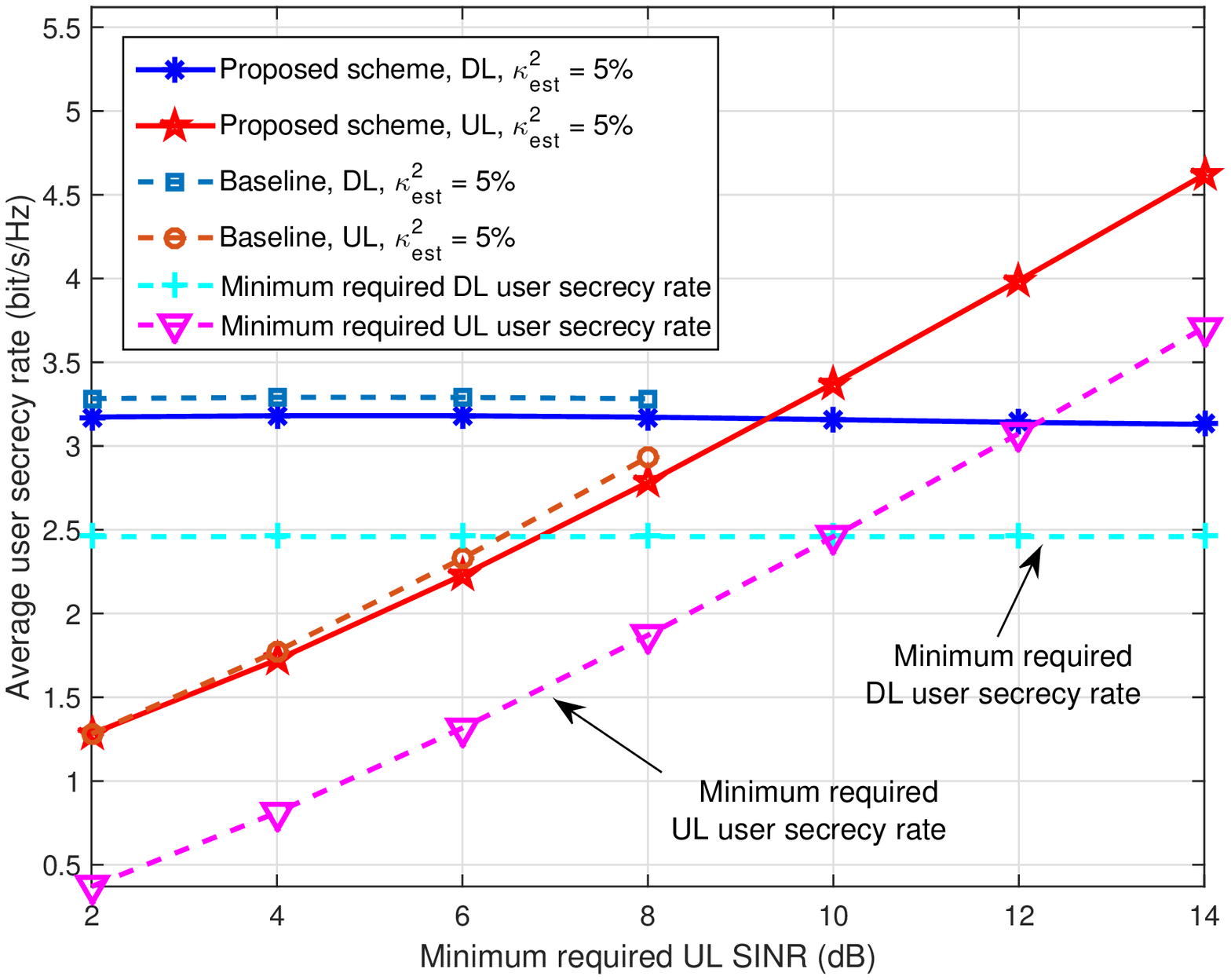}\vspace*{-8mm}
\caption{Average user secrecy rate (bits/s/Hz) versus the minimum required UL SINR (dB), $\Gamma^{\mathrm{UL}}_{\mathrm{req}}$.} \label{fig:SecrecyCap_vs_UL_SINR_err5_10}
 \end{minipage}\vspace*{-8mm}
\end{figure}
Figure \ref{fig:SecrecyCap_vs_DL_SINR_err5_10} depicts the average user secrecy rate of the DL and UL users versus the minimum required DL SINR, $\Gamma^{\mathrm{DL}}_{\mathrm{req}}$, for a maximum normalized channel estimation error of $\kappa_{\mathrm{est}}^2=5\%$ and $\Gamma^{\mathrm{UL}}_{\mathrm{req}}=5$ dB. The FD BS is equipped with $N_\mathrm{T}=10$ antennas. We select the resource allocation policy with $\lambda_1=0.1$ and $\lambda_2=0.9$.
The average user secrecy rates for DL and UL users are calculated by averaging the total DL and the total UL secrecy rates, i.e., $\frac{\sum_{k=1}^K R_{k}^{\mathrm{DL-Sec}}}{K}$ and $\frac{\sum_{j=1}^J R_{j}^{\mathrm{UL-Sec}}}{J}$, respectively.
It can been seen that the average DL user secrecy rate increases with $\Gamma^{\mathrm{DL}}_{\mathrm{req}}$ since the channel capacity between DL user $k$ and potential eavesdropper $m$ is limited to $C_{k,m}^{\mathrm{DL-E}}=1$ bit/s/Hz. Besides, the average UL user secrecy rate depends only weakly on $\Gamma^{\mathrm{DL}}_{\mathrm{req}}$ due to the proposed robust optimization.
In addition, we compare the average DL and UL user secrecy rates of the proposed scheme with the minimum required DL and UL user secrecy rates, i.e., $\log_2(1+\Gamma^{\mathrm{DL}}_{\mathrm{req}}) -\underset{m}{\mathrm{max}}\{R_{\mathrm{tol}_{k,m}}^{\mathrm{DL}}\}$
and
$\log_2(1+\Gamma^{\mathrm{UL}}_{\mathrm{req}})-\underset{m}{\mathrm{max}}\{R_{\mathrm{tol}_{j,m}}^{\mathrm{UL}}\}$, respectively. As can be seen, although the CSI of the DL eavesdropping channels, UL eavesdropping channels, and CCI channels is imperfect, the average user secrecy rate achieved by the proposed scheme fulfills the minimum required user secrecy rate in both DL and UL which confirms that the security of both links can be guaranteed simultaneously. This is due to the robustness of the proposed optimization algorithm design.
On the other hand, the baseline scheme achieves a higher secrecy rate than the proposed scheme for $\Gamma^{\mathrm{DL}}_{\mathrm{req}}$ ranges from $2$ dB to $12$ dB. However, to accomplish this, the baseline scheme requires exceedingly large transmit powers at both the FD BS and the UL users compared to the proposed scheme, cf. Figure \ref{fig:fig_total_power_vs_DL_sinr}. Besides, the superior performance of the baseline scheme in terms of secrecy rate also comes at the expense of a extremely high outage probability. In particular, the baseline scheme is always infeasible when $\Gamma^{\mathrm{DL}}_{\mathrm{req}}$ is larger than $12$ dB, cf. Figure \ref{fig:fig_outage_vs_probability}.

Figure \ref{fig:SecrecyCap_vs_UL_SINR_err5_10} illustrates the average user secrecy rate versus the minimum required UL SINR, $\Gamma^{\mathrm{UL}}_{\mathrm{req}}$, for a maximum normalized channel estimation error of $\kappa_{\mathrm{est}}^2=5\%$ and a minimum required DL SINR of $\Gamma^{\mathrm{DL}}_{\mathrm{req}}=10$ dB. The system setting is the same as for Figure \ref{fig:SecrecyCap_vs_DL_SINR_err5_10}.
It can be seen that the average UL user secrecy rate increases with $\Gamma^{\mathrm{UL}}_{\mathrm{req}}$ since the channel capacity between UL user $j$ and potential eavesdropper $m$ is limited to $C_{j,m}^{\mathrm{UL-E}}=1$ bit/s/Hz. Besides, the average DL user secrecy rate is not sensitive to the increase of $\Gamma^{\mathrm{UL}}_{\mathrm{req}}$.
Similar to Figure \ref{fig:SecrecyCap_vs_DL_SINR_err5_10}, the proposed scheme achieves a higher average user secrecy rate than the minimum required user secrecy rate despite the imperfect CSI which confirms the robustness of the proposed scheme for guaranteeing communication security.
The baseline scheme achieves again a higher average user secrecy rate than the proposed scheme and simultaneous DL and UL secure transmission can be guaranteed when $\Gamma^{\mathrm{UL}}_{\mathrm{req}} \le 8$ dB. However, the baseline scheme consumes significantly larger DL and UL transmit powers than the proposed scheme, cf. Figure \ref{fig:fig_total_power_vs_UL_sinr}. Besides, the baseline scheme incurs again a high outage probability. In fact, the baseline scheme cannot guarantee communication security if $\Gamma^{\mathrm{UL}}_{\mathrm{req}}$ is larger than $8$ dB since it always becomes infeasible.

\begin{figure}[t]
 \centering \vspace*{-10mm}
\includegraphics[width=4in]{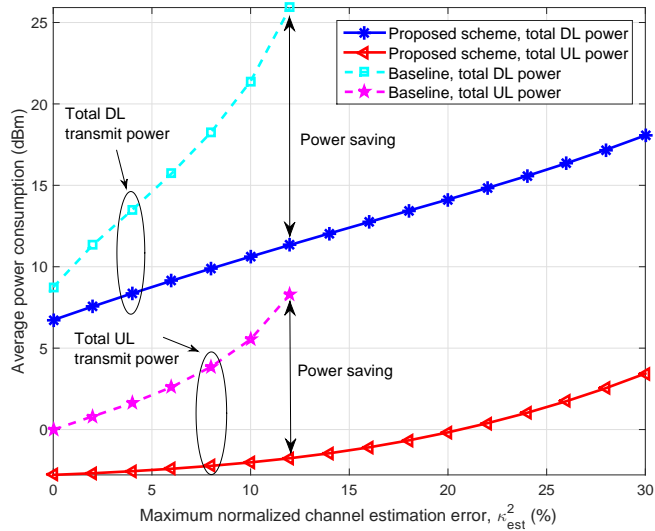} \vspace*{-5mm}
\caption{Average power consumption (dBm) versus the maximum normalized channel estimation error, $\kappa_{\mathrm{est}}^2$.} \label{fig:fig_power_vs_err_nt10_lambda003} \vspace*{-8mm}
\end{figure}

\vspace*{-4mm}
\subsection{Average Transmit Power versus Maximum Channel Estimation Error}
In Figure \ref{fig:fig_power_vs_err_nt10_lambda003}, we investigate the average transmit power consumption versus the maximum normalized channel estimation error, $\kappa_{\mathrm{est}}^2$, for a minimum required DL SINR of $\Gamma^{\mathrm{DL}}_{\mathrm{req}}=10$ dB and a minimum required UL SINR of $\Gamma^{\mathrm{UL}}_{\mathrm{req}}=5$ dB. The FD BS is equipped with $N_\mathrm{T}=10$ antennas. We select again the resource allocation policy with $\lambda_1=0.1$ and $\lambda_2=0.9$ to give preference to the total UL transmit power minimization. As can be observed, the average total DL and UL transmit powers increase with increasing maximum normalized channel estimation error, $\kappa_{\mathrm{est}}^2$. In fact, with increasing imperfectness of the CSI, it is more difficult for the FD BS to perform accurate DL beam steering. Hence, the FD BS transmits the information signal and the AN with higher power to satisfy the DL QoS requirements and to guarantee transmission security in both DL and UL. On the other hand, the higher total DL power causes more SI to the UL reception. As a result, the UL users are forced to transmit with higher powers to satisfy the minimum required receive SINR at the FD BS. As for the baseline scheme, feasible solutions can only be found for $\kappa_{\mathrm{est}}^2 \le 12\%$ which indicates that the maximum channel estimation error tolerance is much lower than for the proposed scheme. Besides, for the baseline scheme, a significantly higher power is consumed for both DL and UL transmission compared to the proposed scheme even if the problem is feasible.

\vspace*{-4mm}
\section{conclusions}
In this paper, we studied the power efficient resource allocation algorithm design for enabling secure MU-MIMO wireless communication with an FD BS. The algorithm design was formulated as a non-convex MOOP via the weighted Tchebycheff method. The proposed problem aimed at jointly minimizing the total DL and UL transmit powers for achieving simultaneous secure DL and UL transmission. The imperfectness of the CSI of the eavesdropping channels and the CCI channels was taken into account for robust resource allocation algorithm design. The proposed MOOP was solved optimally by SDP relaxation. We proved that the globally optimal solution can always be obtained or constructed by solving at most two convex SDP optimization problems. Simulation results not only revealed the trade-off between the total DL and UL transmit power consumption, but also confirm that the proposed FD system provides substantial power savings over a baseline scheme. Furthermore, the simulation results confirmed the robustness of the proposed scheme with respect to imperfect CSI. More importantly, our results revealed that an FD BS can guarantee secure UL transmission which is not possible with an HD BS.

\vspace*{-4mm}
\section*{Appendix}
\subsection{Proof of Proposition 1}
We start the proof by rewriting constraints ${\mbox{C3}}$ and ${\mbox{C4}}$ as \vspace*{-3mm}
\begin{eqnarray}
\label{eqn:det-DL-tol-rate-constraint}\hspace*{-2mm}{\mbox{C3}}\mbox{: }\det(\mathbf{I}_{N_{\mathrm{R}}} \hspace*{-1mm} +\hspace*{-1mm}\mathbf{X}_m^{-1}\mathbf{L}_m^H\mathbf{W}_k\mathbf{L}_m) \hspace*{-1mm} \le \hspace*{-1mm} 2^{R_{\mathrm{tol}_{k,m}}^{\mathrm{DL}}}
\label{eqn:half-det-DL-tol-rate-constraint}\hspace*{-2.5mm}&\overset{(a)}{\Longleftrightarrow}& \hspace*{-2.5mm}\det(\mathbf{I}_{N_{\mathrm{R}}}\hspace*{-1mm} + \hspace*{-1mm} \mathbf{X}_m^{-1/2}\mathbf{L}_m^H\mathbf{W}_k\mathbf{L}_m\mathbf{X}_m^{-1/2}) \hspace*{-1mm} \le \hspace*{-1mm} 2^{R_{\mathrm{tol}_{k,m}}^{\mathrm{DL}}},\\
\label{eqn:det-UL-tol-rate-constraint}\hspace*{-2.5mm}{\mbox{C4}}\mbox{: }\det(\mathbf{I}_{N_{\mathrm{R}}} \hspace*{-1mm} + \hspace*{-1mm} P_j\mathbf{X}_m^{-1}\mathbf{e}_{j,m}\mathbf{e}_{j,m}^H) \hspace*{-1mm} \le \hspace*{-1mm} 2^{R_{\mathrm{tol}_{j,m}}^{\mathrm{UL}}}
\label{eqn:half-det-UL-tol-rate-constraint}\hspace*{-2mm} &\overset{(b)}{\Longleftrightarrow}& \hspace*{-2.5mm} \det(\mathbf{I}_{N_{\mathrm{R}}}\hspace*{-1mm} + \hspace*{-1mm} P_j\mathbf{X}_m^{-1/2}\mathbf{e}_{j,m}\mathbf{e}_{j,m}^H\mathbf{X}_m^{-1/2}) \hspace*{-1mm} \le \hspace*{-1mm}2^{R_{\mathrm{tol}_{j,m}}^{\mathrm{UL}}},
\end{eqnarray}
respectively. $(a)$ and $(b)$ hold due to a basic matrix equality, namely $\det(\mathbf{I}+\mathbf{AB})=\det(\mathbf{I}+\mathbf{BA})$. Then, we study a lower bound of (\ref{eqn:half-det-DL-tol-rate-constraint}) and (\ref{eqn:half-det-UL-tol-rate-constraint}) by applying the following Lemma.

\emph{Lemma 3 \cite{li2013spatially}:}
For any semidefinite matrix $\mathbf{A} \succeq \mathbf{0}$, the inequality $\det(\mathbf{I}+\mathbf{A}) \ge 1+\Tr(\mathbf{A})$ holds where equality holds if and only if $\Rank(\mathbf{A}) \le 1$.

We note that $\mathbf{X}_m^{-1/2}\mathbf{L}_m^H\mathbf{W}_k\mathbf{L}_m\mathbf{X}_m^{-1/2} \succeq \mathbf{0}$ holds in (\ref{eqn:half-det-DL-tol-rate-constraint}). Thus, applying Lemma 3 to (\ref{eqn:half-det-DL-tol-rate-constraint}) yields\vspace*{-2mm}
\begin{eqnarray}\label{eqn:DL-QoS-det-trace}
\det(\mathbf{I}_{N_{\mathrm{R}}}+\mathbf{X}_m^{-1/2}\mathbf{L}_m^H\mathbf{W}_k\mathbf{L}_m\mathbf{X}_m^{-1/2}) \ge 1+\Tr(\mathbf{X}_m^{-1/2}\mathbf{L}_m^H\mathbf{W}_k\mathbf{L}_m\mathbf{X}_m^{-1/2}).
\end{eqnarray}

As a result, by combining (\ref{eqn:det-DL-tol-rate-constraint}) and (\ref{eqn:DL-QoS-det-trace}), we have the following implications\vspace*{-2mm}
\begin{eqnarray}\label{eqn:LMI-DL}
\Tr(\mathbf{X}_m^{-1/2}\mathbf{L}_m^H\mathbf{W}_k\mathbf{L}_m\mathbf{X}_m^{-1/2}) \le \xi_{k,m}^{\mathrm{DL}}
&\overset{(c)}{\Longrightarrow} & \hspace*{-0mm}\lambda_{\mathrm{max}}(\mathbf{X}_m^{-1/2}\mathbf{L}_m^H\mathbf{W}_k\mathbf{L}_m\mathbf{X}_m^{-1/2}) \le \xi_{k,m}^{\mathrm{DL}} \notag\\
\Longleftrightarrow \hspace*{-0mm}\mathbf{X}_m^{-1/2}\mathbf{L}_m^H\mathbf{W}_k\mathbf{L}_m\mathbf{X}_m^{-1/2} \preceq \xi_{k,m}^{\mathrm{DL}}\mathbf{I}_{N_\mathrm{R}}
&\Longleftrightarrow& \hspace*{-0mm}\mathbf{L}_m^H\mathbf{W}_k\mathbf{L}_m \preceq \xi_{k,m}^{\mathrm{DL}}\mathbf{X}_m,
\end{eqnarray}
where $\lambda_{\mathrm{max}}(\mathbf{A})$ denotes the maximum eigenvalue of matrix $\mathbf{A}$ and $(c)$ is due to the fact that $\Tr(\mathbf{A}) \ge \lambda_{\mathrm{max}}(\mathbf{A})$ holds for any $\mathbf{A} \succeq \mathbf{0}$. Besides, if $\Rank(\mathbf{W}_k) \le 1$, we have \vspace*{-2mm}
\begin{eqnarray}
\Rank(\mathbf{X}_m^{-1/2}\mathbf{L}_m^H\mathbf{W}_k\mathbf{L}_m\mathbf{X}_m^{-1/2}) &\le& \min\Big\{\Rank(\mathbf{X}_m^{-1/2}\mathbf{L}_m^H),\Rank(\mathbf{W}_k\mathbf{L}_m\mathbf{X}_m^{-1/2})\Big\} \notag \\
&\le& \Rank(\mathbf{W}_k\mathbf{L}_m\mathbf{X}_m^{-1/2}) \,\,\, \le 1.
\end{eqnarray}
Then, the equality in (\ref{eqn:DL-QoS-det-trace}) holds. Besides, in (\ref{eqn:LMI-DL}),
$\Tr(\mathbf{X}_m^{-1/2}\mathbf{L}_m^H\mathbf{W}_k\mathbf{L}_m\mathbf{X}_m^{-1/2}) \le \xi_{k,m}^{\mathrm{DL}}$ are equivalent with
$\lambda_{\mathrm{max}}(\mathbf{X}_m^{-1/2}\mathbf{L}_m^H\mathbf{W}_k\mathbf{L}_m\mathbf{X}_m^{-1/2}) \le \xi_{k,m}^{\mathrm{DL}}$ .
Therefore, (\ref{eqn:det-DL-tol-rate-constraint}) and (\ref{eqn:LMI-DL}) are equivalent if $\Rank(\mathbf{W}_k) \le 1$.

As for constraint ${\mbox{C4}}$, we note that $\Rank(P_j\mathbf{X}_m^{-1/2}\mathbf{e}_{j,m}\mathbf{e}_{j,m}^H\mathbf{X}_m^{-1/2}) \le 1$ always holds. Therefore, by applying Lemma 3 to (\ref{eqn:half-det-UL-tol-rate-constraint}), we have\vspace*{-2mm}
\begin{eqnarray}
\label{eqn:UL-QoS-det-trace}\det(\mathbf{I}_{N_{\mathrm{R}}}+P_j\mathbf{X}_m^{-1/2}\mathbf{e}_{j,m}\mathbf{e}_{j,m}^H\mathbf{X}_m^{-1/2}) = 1+\Tr(P_j\mathbf{X}_m^{-1/2}\mathbf{e}_{j,m}\mathbf{e}_{j,m}^H\mathbf{X}_m^{-1/2}).
\end{eqnarray}
Then, by combining (\ref{eqn:det-UL-tol-rate-constraint}) and (\ref{eqn:UL-QoS-det-trace}), we have the following implications: \vspace*{-3mm}
\begin{eqnarray}
{\mbox{C4}}\hspace*{-1mm} & \Longleftrightarrow & \hspace*{-2mm} \Tr(P_j\mathbf{X}_m^{-1/2}\mathbf{e}_{j,m}\mathbf{e}_{j,m}^H\mathbf{X}_m^{-1/2}) \le \xi_{j,m}^{\mathrm{UL}}
\Longleftrightarrow  \hspace*{0mm}\lambda_{\mathrm{max}}(P_j\mathbf{X}_m^{-1/2}\mathbf{e}_{j,m}\mathbf{e}_{j,m}^H\mathbf{X}_m^{-1/2}) \le \xi_{j,m}^{\mathrm{UL}}\notag \\
& \Longleftrightarrow & \hspace*{-2mm}P_j\mathbf{X}_m^{-1/2}\mathbf{e}_{j,m}\mathbf{e}_{j,m}^H\mathbf{X}_m^{-1/2} \preceq \xi_{j,m}^{\mathrm{UL}}\mathbf{I}_{N_\mathrm{R}}
\Longleftrightarrow  \hspace*{0mm}P_j\mathbf{e}_{j,m}\mathbf{e}_{j,m}^H \preceq \xi_{j,m}^{\mathrm{UL}}\mathbf{X}_m. \quad
\end{eqnarray}

\vspace*{-9mm}
\subsection{Proof of Proposition 2}
To facilitate the presentation, we define $\mathbf{S}_{j,m}\in\mathbb{H}^{N_{\mathrm{R}}}$ and $\mathbf{T}_{m}\in\mathbb{H}^{N_{\mathrm{R}}}$ as $\mathbf{S}_{j,m}=P_j\mathbf{e}_{j,m}\mathbf{e}_{j,m}^H$ and $\mathbf{T}_{m}=(\xi_{j,m}^{\mathrm{UL}}-1) \mathbf{X}_m$, respectively. If there exists a Hermitian matrix $\mathbf{M}_{j,m} \in \mathbb{H}^{N_{\mathrm{R}}}$ satisfying constraints ${\widetilde{\mbox{C4a}}}$ and ${\widetilde{\mbox{C4b}}}$, then constraints ${\widetilde{\mbox{C4a}}}$ and ${\widetilde{\mbox{C4b}}}$ imply \vspace*{-2mm}
\begin{eqnarray}
\label{eqn:slack_1}
\widetilde{\mbox{C4a}} \Longleftrightarrow \mathbf{x}^H(\mathbf{S}_{j,m}-\mathbf{M}_{j,m})\mathbf{x} \le 0, \text{  and  }\,\,
\label{eqn:slack_2}
\widetilde{\mbox{C4b}} \Longleftrightarrow \mathbf{x}^H(\mathbf{M}_{j,m}-\mathbf{T}_m)\mathbf{x} \le 0,
\end{eqnarray}
where $\mathbf{x} \in \mathbb{C}^{N_{\mathrm{R}}\times 1}$ is any non-zero vector. From (\ref{eqn:slack_2}), we have \vspace*{-4mm}
\begin{eqnarray}
\mathbf{x}^H(\mathbf{S}_{j,m}-\mathbf{T}_m)\mathbf{x} \le 0 \Longleftrightarrow \widetilde{\mbox{C4}}\mbox{: }\mathbf{S}_{j,m} \preceq \mathbf{T}_{m}.
\end{eqnarray}

\vspace*{-13mm}
\subsection{Proof of Theorem 1}
The SDP relaxed version of equivalent Problem $3$ in (\ref{eqn:SDR-MOP-robust}) is jointly convex with respect to the optimization variables and satisfies the Slater's constraint qualification. Therefore, strong duality holds and solving the dual problem is equivalent to solving the primal problem \cite{book:convex}. For obtaining the dual problem, we first need the Lagrangian function of the primal problem in (\ref{eqn:SDR-MOP-robust}) which is given by \vspace*{-2mm}
\begin{eqnarray}
\label{eqn:Lagrangian}\notag{\cal L}&=& \lambda_1\theta_1\sum_{k=1}^{K}\Tr(\mathbf{W}_k)- \sum_{k=1}^{K}\Tr(\mathbf{R}_{\overline{\mathrm{C1}}_k}\big(\mathbf{W}_k,\mathbf{Z},P_j,\delta_k\big)\mathbf{D}_{\overline{\mathrm{C1}}_k})
+\sum_{j=1}^{J}\mu_j\sum_{k=1}^{K}\Tr(\rho\mathbf{W}_k\mathbf{H}_{\mathrm{SI}}^H\mathbf{V}_j\mathbf{H}_{\mathrm{SI}})\\[-2mm]
&-&\sum_{m=1}^{M}\sum_{k=1}^{K}\Tr(\mathbf{R}_{{\overline{\mathrm{C3}}}_{k,m}} \big(\mathbf{W}_k,\mathbf{Z},t_{k,m}\big)\mathbf{D}_{{\overline{\mathrm{C3}}}_{k,m}})- \sum_{k=1}^{K}\Tr(\mathbf{W}_k\mathbf{Y}_k)+\Delta.
\end{eqnarray}
Here, $\Delta$ denotes the collection of terms that only involve variables that are independent of $\mathbf{W}_k$. $\mu_j$ and $\theta_i$ are the Lagrange multipliers associated with constraints ${\mbox{C2}}$ and ${\mbox{C9}}$, respectively. Matrices $\mathbf{D}_{\overline{\mathrm{C1}}_k} \in {\mathbb{C}^{{(J+1)}\times {(J+1)}}}$ and $\mathbf{D}_{{\overline{\mathrm{C3}}}_{k,m}} \in {\mathbb{C}^{(N_{\mathrm{T}}+N_{\mathrm{R}})\times (N_{\mathrm{T}}+N_{\mathrm{R}})}} $ are the Lagrange multiplier matrices for constraints $\overline{\text{C1}}$ and $\overline{\text{C3}}$, respectively. Matrix $\mathbf{Y}_k \in {\mathbb{C}^{N_{\mathrm{T}}\times N_{\mathrm{T}}}} $ is the Lagrange multiplier matrix for the positive semidefinite constraint ${\text{C7}}$ on matrix $\mathbf{W}_k$. For notational simplicity, we define $\Psi$ as the set of scalar Lagrange multipliers for constraints $\mbox{C5}$, $\mbox{C9}$, and $\mbox{C10}$ and $\mathbf{\Phi}$ as the set of matrix Lagrange multipliers for constraints $\overline{\mbox{C1}}$, $\mbox{C2}$, $\overline{\mbox{C3}}$, $\overline{\mbox{C4}}\mbox{a}$, $\overline{\mbox{C4}}\mbox{b}$, $\mbox{C6}$, and $\mbox{C7}$.
Thus, the dual problem for the SDP relaxed problem in (\ref{eqn:SDR-MOP-robust}) is given by \vspace*{-3mm}
\begin{eqnarray}\label{eqn:dual-constraint}
\underset{\Psi \ge 0, \mathbf{\Phi} \succeq \mathbf{0}}{\maxo} \,\,\underset{\underset{ P_j,\tau,\delta_k, t_{k,m},\alpha_{j,m},\beta_{j,m}}{\mathbf{W}_k,\mathbf{Z}\in\mathbb{H}^{N_{\mathrm{T}}},\mathbf{M}_{j,m}\in\mathbb{H}^{N_{\mathrm{R}}},}}{\mino} \,\,{\cal L} \Big(\hspace*{-0.5mm}\mathbf{W}_k,\mathbf{Z},\mathbf{M}_{j,m},P_j,\tau,\delta_k, t_{k,m},\alpha_{j,m},\beta_{j,m},\Psi, \mathbf{\Phi}\hspace*{-0.5mm}\Big)\notag\\[-7mm]
&&\hspace*{-90mm}\mbox{s.t.}\hspace*{3mm} \sum_{i=1}^2 \theta_i = 1.
\end{eqnarray}
Constraint $\sum_{i=1}^2 \theta_i = 1$ is imposed to guarantee a bounded solution of the dual problem \cite{book:convex}. Then, we reveal the structure of the optimal $\mathbf{W}_k$ of (\ref{eqn:SDR-MOP-robust}) by studying the Karush-Kuhn-Tucker (KKT) conditions.
The KKT conditions for the optimal $\mathbf{W}_k^*$ are given by:\vspace*{-3mm}
\begin{eqnarray}\label{eqn:KKT-multiplier}\mathbf{Y}_k^*,\mathbf{D}_{\overline{\mathrm{C1}}_k}^*,\mathbf{D}_{{\overline{\mathrm{C3}}}_{k,m}}^*
\hspace*{-3mm}&\succeq&\hspace*{-3mm} \mathbf{0},\quad\mu_j^*, \theta_i^* \ge 0,\\[-1mm]
 \mathbf{Y}_k^*\mathbf{W}_k^*\hspace*{-3mm}&=&\hspace*{-3mm}\mathbf{0}, \label{eqn:KKT-complementarity}\\[-2mm]
\nabla_{\mathbf{W}_k^*}{\cal L}\hspace*{-3mm}&=&\hspace*{-3mm} \mathbf{0}, \label{eqn:subgradient}
\end{eqnarray}
where $\mathbf{Y}_k^*$, $\mathbf{D}_{\overline{\mathrm{C1}}_k}^*$, $\mathbf{D}_{{\overline{\mathrm{C3}}}_{k,m}}^*$, $\mu_j^*,$ and $\theta_i^*$ are the optimal Lagrange multipliers for dual problem (\ref{eqn:dual-constraint}), $\nabla_{\mathbf{W}_k^*}{\cal L}$ denotes the gradient of Lagrangian function ${\cal L}$ with respect to matrix $\mathbf{W}_k^*$. The KKT condition in (\ref{eqn:subgradient}) can be expressed as \vspace*{-1mm}
\begin{eqnarray}\label{eqn:KKT-gradient-equivalent}
\lambda_1\theta_1\mathbf{I}_{N_{\mathrm{T}}}+ \sum_{j=1}^{J}\mu_j^*\rho\mathbf{H}_{\mathrm{SI}}^H\diag(\mathbf{V}_j)\mathbf{H}_{\mathrm{SI}}+ \sum_{m=1}^{M}\mathbf{B}_{\mathbf{L}_m}\frac{\mathbf{D}_{{\overline{\mathrm{C3}}}_{k,m}}^*} {\Gamma^{\mathrm{DL}}_{\mathrm{req}_k}}\mathbf{B}_{\mathbf{L}_m}^H =\mathbf{Y}_k^*+ \mathbf{B}_{\mathbf{h}_k}\frac{\mathbf{D}_{\overline{\mathrm{C1}}_k}^*}{\Gamma^{\mathrm{DL}}_{\mathrm{req}_k}}\mathbf{B}_{\mathbf{h}_k}^H.
\end{eqnarray}
Now, we divide the proof into two cases according to the value of $\lambda_1$. First, for the case of $0 < \lambda_1 \le 1$, we define \vspace*{-2mm}
\begin{eqnarray}\label{eqn:KKT-mess-B}
\mathbf{A}_k^*=\sum_{j=1}^{J}\mu_j^*\rho\mathbf{H}_{\mathrm{SI}}^H\diag(\mathbf{V}_j)\mathbf{H}_{\mathrm{SI}}+ \sum_{m=1}^{M}\mathbf{B}_{\mathbf{L}_m}\frac{\mathbf{D}_{{\overline{\mathrm{C3}}}_{k,m}}^*} {\Gamma^{\mathrm{DL}}_{\mathrm{req}_k}}\mathbf{B}_{\mathbf{L}_m}^H, \text{  and} \,\,\,
\mathbf{\Pi}_k^*=\lambda_1\theta_1\mathbf{I}_{N_{\mathrm{T}}}+\mathbf{A}_k^*,
\end{eqnarray}
for notational simplicity. Then, (\ref{eqn:KKT-gradient-equivalent}) implies \vspace*{-5mm}
\begin{eqnarray}\label{eqn:KKT-shorter}
\mathbf{Y}^*=\mathbf{\Pi}_k^*-\mathbf{B}_{\mathbf{h}_k}\frac{\mathbf{D}_{\overline{\mathrm{C1}}_k}^*} {\Gamma^{\mathrm{DL}}_{\mathrm{req}_k}}\mathbf{B}_{\mathbf{h}_k}^H.
\end{eqnarray}
Premultiplying both sides of (\ref{eqn:KKT-shorter}) by $\mathbf{W}_k^*$, and utilizing (\ref{eqn:KKT-complementarity}), we have \vspace*{-0.7mm}
\begin{eqnarray}\label{eqn:KKT-premultiply}
\mathbf{W}_k^*\mathbf{\Pi}_k^* =\mathbf{W}_k^*\mathbf{B}_{\mathbf{h}_k}\frac{\mathbf{D}_{\overline{\mathrm{C1}}_k}^*}{\Gamma^{\mathrm{DL}}_{\mathrm{req}_k}}\mathbf{B}_{\mathbf{h}_k}^H.
\end{eqnarray}
By applying basic inequalities for the rank of matrices, the following relation holds:
\begin{eqnarray}\label{eqn:rank-Wk}
\Rank(\mathbf{W}_k^*)\hspace*{-1mm}&\overset{(a)}{=}&\hspace*{-1mm}\Rank(\mathbf{W}_k^*\mathbf{\Pi}_k^*)\notag =\Rank\Big(\mathbf{W}_k^*\mathbf{B}_{\mathbf{h}_k}\frac{\mathbf{D}_{\overline{\mathrm{C1}}_k}^*} {\Gamma^{\mathrm{DL}}_{\mathrm{req}_k}}\mathbf{B}_{\mathbf{h}_k}^H\Big)\notag\\[-2mm]
&\overset{(b)}{\le}&\hspace*{-1mm}\min\Big\{\Rank(\mathbf{W}_k^*), \Rank\Big(\mathbf{B}_{\mathbf{h}_k}\frac{\mathbf{D}_{\overline{\mathrm{C1}}_k}^*} {\Gamma^{\mathrm{DL}}_{\mathrm{req}_k}}\mathbf{B}_{\mathbf{h}_k}^H\Big)\Big\} \overset{(c)}{\le} \Rank\Big(\mathbf{B}_{\mathbf{h}_k}\frac{\mathbf{D}_{\overline{\mathrm{C1}}_k}^*} {\Gamma^{\mathrm{DL}}_{\mathrm{req}_k}}\mathbf{B}_{\mathbf{h}_k}^H\Big),
\end{eqnarray}
where $(a)$ is due to $\mathbf{\Pi}_k^* \succ \mathbf{0}$, $(b)$ is due to the basic result $\Rank(\mathbf{A}\mathbf{B}) \le \min \big\{ \Rank(\mathbf{A}),\Rank(\mathbf{B})\big\}$, and $(c)$ is due to the fact that $\min\{a,b\} \le a$. In order to further reveal the structure of $\mathbf{W}_k^*$ in (\ref{eqn:rank-Wk}), we study the rank of $\mathbf{B}_{\mathbf{h}_k}\frac{\mathbf{D}_{\overline{\mathrm{C1}}_k}^*}{\Gamma^{\mathrm{DL}}_{\mathrm{req}_k}}\mathbf{B}_{\mathbf{h}_k}^H$ which is given by \vspace*{-2mm}
\begin{eqnarray}\label{eqn:rank-multiplier-C1-QC}
&&\Rank\Big(\mathbf{B}_{\mathbf{h}_k}\frac{\mathbf{D}_{\overline{\mathrm{C1}}_k}^*}{\Gamma^{\mathrm{DL}}_{\mathrm{req}_k}}\mathbf{B}_{\mathbf{h}_k}^H\Big) =\Rank\Big(\big[\mathbf{0}\quad\mathbf{h}_k\big]\frac{\mathbf{D}_{\overline{\mathrm{C1}}_k}^*} {\Gamma^{\mathrm{DL}}_{\mathrm{req}_k}}\mathbf{B}_{\mathbf{h}_k}^H\Big)\notag\\[-3mm]
&\overset{(d)}{\le}&\min\Big\{\Rank\Big(\big[\mathbf{0}\quad\mathbf{h}_k\big]\Big), \Rank\Big(\frac{\mathbf{D}_{\overline{\mathrm{C1}}_k}^*} {\Gamma^{\mathrm{DL}}_{\mathrm{req}_k}}\mathbf{B}_{\mathbf{h}_k}^H\Big)\Big\}\overset{(e)}{\le}\Rank\Big(\big[\mathbf{0}\quad\mathbf{h}_k\big]\Big) \le 1,
\end{eqnarray}
where for $(d)$ and $(e)$, we used the same results as for $(b)$ and $(c)$, respectively. By combining (\ref{eqn:rank-Wk}) and (\ref{eqn:rank-multiplier-C1-QC}), the rank of $\mathbf{W}_k^*$ is given by \vspace*{-4mm}
\begin{eqnarray}\label{eqn:rank-Wk-inequality}
\Rank(\mathbf{W}_k^*)&\le&\Rank\Big(\mathbf{B}_{\mathbf{h}_k}\frac{\mathbf{D}_{\overline{\mathrm{C1}}_k}^*} {\Gamma^{\mathrm{DL}}_{\mathrm{req}_k}}\mathbf{B}_{\mathbf{h}_k}^H\Big) \le 1.
\end{eqnarray}
We note that $\mathbf{W}_k^* \neq \mathbf{0}$ for $\Gamma^{\mathrm{DL}}_{\mathrm{req}_k} > 0$. Thus, $\Rank(\mathbf{W}_k^*)=1$.

Then, for the case of $\lambda_1=0$, we show that we can always construct a rank-one optimal solution $\mathbf{W}_k^{**}$. We note that the problem in (\ref{eqn:SDR-MOP-robust}) with $\lambda_1=0$ is equivalent to a total UL transmit power minimization problem which is given by \vspace*{-7mm}
\begin{eqnarray}
\label{eqn:lambda1-0-ulprob}&&\hspace*{0mm}\underset{P_j,\delta_k, t_{k,m},\alpha_{j,m},\beta_{j,m}} {\underset{\mathbf{W}_k,\mathbf{Z}\in\mathbb{H}^{N_{\mathrm{T}}}, \mathbf{M}_{j,m}\in\mathbb{H}^{N_{\mathrm{R}}},} {\mino}}\,\, \sum_{j=1}^{J}P_j \notag\\
\mbox{s.t.} &&\hspace*{4mm}\overline{\mbox{C1}},\mbox{C2},\overline{\mbox{C3}},\overline{\mbox{C4}},\mbox{C5},\mbox{C6},\mbox{C7},{\mbox{C9}},{\mbox{C10}}.
\end{eqnarray}
We first solve the above convex optimization problem and obtain the UL transmit power $P_j^{**}$, the DL beamformimg matrix $\mathbf{W}_k^*$, the AN covariance matrix $\mathbf{Z}^*$, and the optimal auxiliary variables which are collected in $\mathbf{\Xi}^* \triangleq \{ \mathbf{M}_{j,m}^*,\delta_k^*, t_{k,m}^*,\alpha_{j,m}^*,\beta_{j,m}^*\}$. If $\Rank(\mathbf{W}_k^*)=1,\forall k$, then the globally optimal solution of problem \eqref{eqn:MOP-transformed} for $\lambda_1=0$ is achieved. Otherwise, we substitute $P_j^{**}$, $\mathbf{Z}^*$, and $\mathbf{\Xi}^*$ into the following auxiliary problem: \vspace*{-3mm}
\begin{eqnarray}
\label{eqn:lambda1-0-dlprob}\underset{} {\underset{\mathbf{W}_k \in \mathbb{H}^{N_{\mathrm{T}}}}{\mino}}\,\, \hspace*{-7mm}&&\hspace*{0mm}\sum_{k=1}^{K}\Tr(\mathbf{W}_k)+\Tr(\mathbf{Z}^*) \notag\\[-2mm]
\mbox{s.t.} \hspace*{0mm}&&\hspace*{0mm}\overline{\mbox{C1}},\mbox{C2},\overline{\mbox{C3}},\overline{\mbox{C4}},\mbox{C5},\mbox{C6},\mbox{C7},{\mbox{C9}},{\mbox{C10}}.
\end{eqnarray}
Since the problem in (\ref{eqn:lambda1-0-dlprob}) shares the feasible set of problem (\ref{eqn:lambda1-0-ulprob}), problem (\ref{eqn:lambda1-0-dlprob}) is also feasible. Now, we claim that for a given $P_j^{**}$, $\mathbf{Z}^*$, and $\mathbf{\Xi}^*$ in (\ref{eqn:lambda1-0-dlprob}), the solution $\mathbf{W}_k^{**}$ of (\ref{eqn:lambda1-0-dlprob}) is a rank-one matrix. First, the gradient of the Lagrangian function for (\ref{eqn:lambda1-0-dlprob}) with respect to $\mathbf{W}_k^{**}$ can be expressed as \vspace*{-7mm}
\begin{eqnarray}\label{eqn:KKT-shorter-dl}
\mathbf{Y}^{**}=\mathbf{\Pi}_k^{**}-\mathbf{B}_{\mathbf{h}_k}\frac{\mathbf{D}_{\overline{\mathrm{C1}}_k}^{**}} {\Gamma^{\mathrm{DL}}_{\mathrm{req}_k}}\mathbf{B}_{\mathbf{h}_k}^H,
\end{eqnarray}
where $\mathbf{\Pi}_k^{**}=\mathbf{I}_{N_{\mathrm{T}}}+\mathbf{A}_k^{**}$ and $\mathbf{A}_k^{**}=\sum_{j=1}^{J}\mu_j^{**}\rho\mathbf{H}_{\mathrm{SI}}^H\diag(\mathbf{V}_j)\mathbf{H}_{\mathrm{SI}}+ \sum_{m=1}^{M}\mathbf{B}_{\mathbf{L}_m}\frac{\mathbf{D}_{{\overline{\mathrm{C3}}}_{k,m}}^{**}} {\Gamma^{\mathrm{DL}}_{\mathrm{req}_k}}\mathbf{B}_{\mathbf{L}_m}^H$, and $\mathbf{Y}_k^{**},\mathbf{D}_{\overline{\mathrm{C1}}_k}^{**}$, $\mathbf{D}_{{\overline{\mathrm{C3}}}_{k,m}}^{**}$, and $\mu_j^{**}$ are the optimal Lagrange multipliers for the dual problem of (\ref{eqn:lambda1-0-dlprob}).
Premultiplying both sides of (\ref{eqn:KKT-shorter-dl}) by the optimal solution $\mathbf{W}_k^{**}$, we have \vspace*{-2mm}
\begin{eqnarray}\label{eqn:KKT-premultiply-2}
\mathbf{W}_k^{**}\mathbf{\Pi}_k^{**} =\mathbf{W}_k^{**}\mathbf{B}_{\mathbf{h}_k}\frac{\mathbf{D}_{\overline{\mathrm{C1}}_k}^{**}} {\Gamma^{\mathrm{DL}}_{\mathrm{req}_k}}\mathbf{B}_{\mathbf{h}_k}^H.
\end{eqnarray}
We note that $\mathbf{\Pi}_k^{**}$ is a full-rank matrix, i.e., $\mathbf{\Pi}_k^{**} \succ \mathbf{0}$, and (\ref{eqn:KKT-premultiply-2}) has the same form as (\ref{eqn:KKT-premultiply}). Thus, we can follow the same approach as for the case of  $0<\lambda_i\leq 1$ for showing that $\mathbf{W}_k^{**}$ is a rank-one matrix. Also, since $\mathbf{W}_k^{**}$ is a feasible solution of (\ref{eqn:lambda1-0-ulprob}) for $P_j^{**}$, an optimal rank-one matrix $\mathbf{W}_k^{**}$ for the case of $\lambda_1 = 0$ is constructed.
\vspace*{-3mm}


\end{document}